\documentclass[12pt]{article}
\usepackage{enumerate}
\usepackage{amsfonts}
\usepackage{amssymb}
\usepackage{onlyamsmath}
\usepackage[OT4]{fontenc}
\usepackage[cp1250]{inputenc}
\title{Global Space-time Symmetries of Quantized Euclidean and Minkowski Superspaces}
\author{ C. Gonera\footnote{e-mail: cgonera@uni.lodz.pl}  \quad M. Wodzisławski
\\ \\
Department of Theoretical Physics and Computer Science,\\
University of Łódź,\\
Pomorska 149/153, 90-236 Łódź, Poland }
\date{}

\begin{document}
\maketitle
\begin{abstract}
 Starting with assumptions both simple and natural from "physical" point of view we present a direct construction of the transformations preserving
wide class of (anti)commutation relations which describe Euclidean/Minkowski superspace quantizations. These generalized transformations act on deformed superspaces as the ordinary ones do on undeformed spaces but they depend on non(anti)commuting parameters satisfying some consistent (anti)commutation relations. Once the coalgebraic structure compatible with the algebraic one is introduced in the set of transformations we deal
with quantum symmetry supergroup. This is the case for intensively studied so called $N = \frac{1}{2}$ supersymmetry as well as its three parameter
extension. The resulting symmetry transformations - supersymmetric extension of $\theta$ - Euclidean group can be regarded as global counterpart of
appropriately twisted Euclidean superalgebra that has been shown to preserve $N = \frac{1}{2}$ supersymmetry.     
   
\end{abstract}
\section{Introduction}
Practically, any reasonable physical theory refers, in more or less explicit way, to assumptions concerning space-time structure and its symmetries.
Usually, it is postulated that space-time has a structure of smooth manifold parameterized (in general, locally) by four real commuting coordinates.
This does not lead to any essential controversy in classical or nonrelativistic quantum theory. However, the validity of this assumption is not
evident when trying to combine quantum theory with special or general theory of relativity in a consistent way. Indeed, in relativistic quantum
theories a possibility of probing the space-time is limited by occurrence of creation/annihilation processes or even gravitational collapse (when
gravitation is taken into account ) if energy density contained in a tiny volume is large enough to create  new particles  or to form a black hole.
Therefore, it is believed that at short-distance scale the model of smooth and commutative space-time should be somehow modified or even replaced by 
some other mathematical framework.

In the simplest scenario, instead of local coordinates $ x^m $, $ m = 0,1,2,3 $, hermitian generators $ \hat x^m $are introduced. They obey commutation
relations 
 
\begin{align}
[\hat{x}^{m}, \hat{x}^{n}]=i\Theta^{mn}
\label{1}
\end{align}

where $\Theta^{mn}$ is real antisymmetric (in general, $x^m$ - depending ) matrix of dimension $ (length)^2 $ 
\cite{a2}, \cite{a3}, \cite{a4}, \cite{a5}, \cite{a6}, \cite{a6.1}, \cite{a7}, \cite{a8}, \cite{a8.1}, \cite{a8.2}.

For example, in the case of so called $\theta $ - Minkowski space this matrix is constant and plays a role analogous to Planck's constant. In
particular, there are space-time uncertainty relations $\Delta x^i\Delta x^j \geq \frac{1}{2}\mid \Theta^{ij}\mid $ giving a bound to the 
resolution in which space-time itself can be probed. As a consequence there is no definite notion of a "point".

These simple semi-classical arguments demonstrate that a sort of space-time quantization is expected to be a generic feature of quantum theory
consistent with relativity.

Although the idea of noncommutative space-time is almost as old as quantum mechanics is  \cite{a9} a real impetus to the study of the subject has been 
provided  by string theory \cite{a10}, \cite{a10.1}, \cite{a10.2}, \cite{a10.3}, \cite{a10.4}, \cite{a10.5}, \cite{a11}, \cite{a11.1}, \cite{a11.2}. It has appeared that if the endpoints of open strings are confined to propagate  on a D-brane in a constant B-field background, then they live effectively on 
noncommutative space whose coordinates satisfy the commutation relations given by eq.(\ref{1}).

It is worth noting here that noncommutative spaces can be regarded as usual ones equipped, however, with additional structure so called 
star product \cite{a12}, \cite{a13}, \cite{a14}, \cite{a15}, \cite{a15.1}, \cite{a16}, \cite{a17}.
For instance, in the case of $\theta $ - Minkowski space an appropriate star product of two functions on this space is 
given by the equation   

\begin{align}\label{2}
&f(x) \star g(x) =\mu(exp\{\frac{1}{2}\Theta^{mn}\partial_{m} \otimes \partial_{n}\}f \otimes g)(x)\\
&\mu(f \otimes g) = fg \nonumber
\end{align}

Now, given deformed space-time one is faced with the important question of its symmetries. The point is that, in general, the 
commutation relations defining space deformation break its standard (i.e. the ones corresponding to undeformed case)
symmetries \cite{a21}, \cite{a22}, \cite{a23}, \cite{a24}. 
For example, the relations given by  eq.(\ref{1}) violate Lorentz invariance (if Lorentz transformations are assumed to be "undeformed" and
to act in the usual way).

This fundamental space-time symmetries breaking is one of the major problems around theories on noncommutative spaces.
There are at least two possible approaches to this question.\\
In the first one, $\Theta^{mn}$ matrix is assumed to distinguish a reference frame and  symmetry transformation group is restricted to a stability group of this matrix. It appears that the stability group ( $SO(1,1) \times SO(2)$
for time noncommutativity or $O(1,1) \times O(2)$ in the case of space noncommutativity \cite{a25}, \cite{a26} ) being abelian one has only one-dimensional irreducible representations. Consequently, in addition to violation of relativity principle the notion of spin becomes unclear in theories with such symmetry group.\\
In the second approach, proceeding  in the spirit of  quantum group theory  \cite{a27}, \cite{a28} \cite{a28.1}, \cite{a29}, \cite{a30}
 deformed algebras and group transformations which leave relations defining noncommutative space unchanged
 \cite{a31}, \cite{a32}, \cite{a32.1}, \cite{a33}, \cite{a34}, \cite{a36} are introduced.

In the case of $\theta $ - Minkowski space such deformed algebra, so called twisted Poincare algebra, has been introduced in Ref.\cite{a32.1}. A starting point there was an observation that the differential operator defining the star product given by eq.(\ref{2}) 
can be interpreted as the inverse of the so called twist operator F \cite{a37}. The star product given by the twist operator     
is covariant with respect to the twisted Poincare algebra. The algebraic sector of this algebra coincides with the usual Poincare algebra one.
Consequently, the representation content of the twisted symmetry algebra is the same as the nontwisted one. That is the main advantage of this
approach justifying the use of the ordinary representations of Poincare algebra in the study of theories on $\theta $ - Minkowski space even though
the ordinary Poincare algebra is not the symmetry algebra of these theories.\\
Having the twist operator one can construct first a universal R matrix and then, referring to Faddeev - Reshetikhin -Takhtajan (FRT) \cite{a38} method for example, the global counterpart of twisted Poincare algebra - so called $\theta $- Poincare group \cite{a32}, \cite{a40}, which is in fact noncommutative Hopf algebra of functions on Poincare group. The $\theta $ -Poincare group can also be obtained in more direct way starting with the following assumptions:
\begin{enumerate}[a)]
\item $\Theta$-Poincare group acts on $\Theta$-Minkowski space as the usual Poincare group does on undeformed Minkowski space. 
\item $\Theta$ -Poincare group transformations commute with Lorentz transformations as well as with $\Theta$ - Minkowski space coordinates.
\item Commutation relations defining $\theta $ - Minkowski space are preserved under $\Theta$ -Poincare group action.
\end{enumerate}    
This, based on deformed algebras/group, approach to symmetries of noncommutative Minkowski space can be applied without essential modifications in
Euclidean case. The application of this scheme to symmetries of non(anti)commutative supersymmetric extensions of Euclidean/Minkowski space
\cite{a45}, \cite{a46}, \cite{a47}, \cite{a48}, \cite{a50}, \cite{a51}, \cite{a52}, \cite{a53}, \cite{a54} is more 
involved. A new ingredient here is the possibility of additional non-trivial contributions to deformations of Euclidean/Minkowski (super)-space
stemming from deformations of grassmannian sector of superspace.\\
Twisted superalgebra technique has been used to study various deformations of Euclidean superspace in Refs.
\cite{a55}, \cite{a56}, \cite{a57}, \cite{a58}, \cite{a59}, \cite{a60}, \cite{a61}. Some preliminary description
concerning only one type of deformation of Euclidean superspace in terms of deformed Euclidean supergroup (global counterpart of twisted superalgebra)
can be found in Ref.\cite{a36}.\\
The present paper and the companion one are devoted to more systematic study of supersymmetric extensions of $\theta $ - Poincare group
and its Euclidean counterpart - $\theta $-Euclidean group as well as to the construction of covariant ( with respect to these extensions ) (anti)commutation relations defining deformations of relevant superspaces.\\
In the first paper we start with simple, intuitive and natural from "physical" point of view assumptions basically concerning the action of generalized Euclidean/Poincare transformations on corresponding deformed superspace and proceed in a spirit of Corrigan et al., Manin, Takhtajan, Wess, Zumino approach to (super)space deformations \cite{a62}, \cite{a63}, \cite{a64}, \cite{a65}. 
This leads to general relations between (anti)commmutation rules of deformed space-time coordinates and 
(anti)commmutation ones of generalized Euclidean/Poincare transformation parameters. Then taking into account the structure of these relations and
requiring the preservation of commutators defining space-time deformation under these generalized transformations allows one to find (for a given consistent grassmannian superspace sector deformation) a sufficient conditions on (anti)commutation rules of deformed superspace coordinates as well as on commutators of relevant transformation parameters.\\
In a case of probably the simplest grassmannian sector superspace deformation given by constant matrices our procedure gives algebraic sector of global symmetries of wide class of deformed Euclidean/Minkowski superspace. In fact, most of the deformations of this type discussed in literature \cite{a36}, \cite{a55}, \cite{a56}, \cite{a57}, \cite{a58}, \cite{a59}, belong to our class. 
In general, (anti)commutation relations obtained in this way do not satisfy Jacobi identities. Imposing these identities gives some constraints on matrices defining deformations. It appears that in Minkowski superspace case Jacobi identities can be satisfied provided the matrix elements of these matrices belong to some Grassmann algebra. Finally, the construction of algebraic sector of symmetry transformations is completed by introducing coalgebraic structure given by appropriately defined coproduct, counit and antipode maps. If it happens that both structure algebraic and coalgebraic ones are consistent one is dealing with non(anti)commutative Hopf superalgebras i.e. quantum supergroups. Such symmetry transformations are analyzed within the so called R - matrix  (or FRT) approach as well as in the framework of star product in the companion paper \cite{a66}. It appears that all three methods lead directly to the same results.\\
In order to make the paper more readable we will generally omit the prefix "anti" and use the words "commutators", "commutation rules", etc. irrespectively of the parity of variables under consideration. However, we keep the standard notation for the commutators ($[. , .]$) and the anticommutators  ($\{. , .\}$).

\section{Global Symmetries of Deformed Superspaces}
\sloppy
 
\subsection{N=1 deformed Euclidean Superspace }

Roughly speaking, (super)space deformations as well as construction of their symmetries can be considered as a procedure of replacing 
commuting coordinates of undeformed (super)spaces and commuting parameters of transformations acting on these coordinates 
by noncommuting quantities satisfying some well-defined, consistent commutation rules. In most cases, this vague definition can be
made mathematically sound within framework of non(anti)commutative Hopf (super)algebras theory.
Nevertheless, we will follow this slightly informal approach believing that it can be formalized if needed.

So we start with simplest $N = 1$ Euclidean superspace \cite{a41}, \cite{a42}, \cite{a43}, \cite{a44} which is parametrized by 
four commuting real coordinates $x^{m}$ $(m=0,1,2,3)$ and 
by four anticommuting grassmannian variables $\eta^{\alpha}$, $ \bar{\eta}^{\dot{\alpha}}$ ($\alpha=1,2$, $\dot{\alpha}=\dot{1},\dot{2})$.
The metric structure in bosonic sector is given by metric tensor \\
$g^{mn}=g_{mn} \equiv diag(1,1,1,1) = \delta_{mn}$ i.e. $x^{m}=x_{m} =g_{mn}x^{m}$. 
In grassmannian sector the role of $g^{mn}$ is played by completely antisymmetric symbols    

\begin{align}\label{3}
\tilde{\varepsilon}^{\alpha \beta} =  \tilde{\varepsilon}^{\dot{\alpha} \dot{\beta}} =\begin{bmatrix}0&&-1\\1&&0\end{bmatrix}\\
\varepsilon_{\alpha \beta} =  \varepsilon_{\dot{\alpha} \dot{\beta}} =\begin{bmatrix}0&&1\\-1&&0\end{bmatrix}\label{4}
\end{align}

\begin{align}\label{5}
\eta^{\alpha} = \tilde{\varepsilon}^{\alpha \beta}\eta_{\beta} \; , \; \bar{\eta}^{\dot{\alpha}} = \tilde{\varepsilon}^{\dot{\alpha} 
\dot{\beta}}\bar{\eta}_{\dot{\beta}}\nonumber \\
\eta_{\alpha} = \varepsilon_{\alpha \beta}\eta^{\beta} \; , \; \bar{\eta}_{\dot{\alpha}} = \varepsilon_{\dot{\alpha} 
\dot{\beta}}\bar{\eta}^{\dot{\beta}}
\end{align}

This superspace can be regarded as the coset one resulting from dividing $N = 1$ Euclidean supergroup by group of four dimensional rotations.\\
Let $M_{mn}$ and $P_{m}$ ($ m, n = 0, 1, 2, 3$ ) be the generators of rotations $( \omega^{mn}=-\omega^{nm} )$ respectively translations $(a^{m})$ in 
$\mathbb{R}^{4}$ while $Q_{\alpha}, \bar{Q}_{\dot{\alpha}}$  ($\alpha$=1, 2,  $\dot{\alpha}\,=\,\dot{1},\dot{2}$) - the additional odd generators of 
supertranslations ($\xi^{\alpha}, \bar{\xi}^{\dot{\alpha}}$).\\
A generic element of $N = 1$ Euclidean supergroup can be written in the following exponential parametrization 

\begin{align}\label{6}
g(\xi, \bar{\xi}, a, \omega) &=exp\left\{i(\xi^{\alpha}Q_{\alpha} + \bar{Q}_{\dot{\alpha}}\bar{\xi}^{\dot{\alpha}})\right\}
exp\left\{ia^{m}P_{m}\right\}exp\left\{\frac{i}{2}\omega^{mn}M_{mn} \right\}
\end{align}

Let us remind the standard notation ( below $\sigma_{i},\  \  i=1,2,3$, are Pauli matrices )

\begin{align} \label{7}
&\sigma_{m} =(iI,\sigma_{i})\nonumber\\
&\bar{\sigma}_{m} =(iI, -\sigma_{i})\nonumber \\
\end{align}

\begin{align}\label{8}
\sigma_{mn} = \frac{i}{4}(\sigma_{m}\bar{\sigma}_{n}-\sigma_{n}\bar{\sigma}_{m})\nonumber \\
\bar{\sigma}_{mn} = \frac{i}{4}(\bar{\sigma}_{m}\sigma_{n}-\bar{\sigma}_{n}\sigma_{m}).\nonumber\\ 
\end{align}.

$\sigma_{mn}$ ( resp. $\bar{\sigma}_{mn}$ ) are generators of $D^{(\frac{1}{2},0)}$ (resp. $ D^{(0,\frac{1}{2})}$) representations of 
$sO(4)$ $\sim $ $su(2)\bigoplus su(2)$ algebra.\\

The generators $Q_{\alpha}$, $\bar{Q}_{\dot{\alpha}}$, $P_{m}$, $M_{mn}$ satisfy commutation relations defining $N=1$ Euclidean superalgebra:\\

\begin{align}\label{9} 
&[M_{ab}, M_{cd}] = -i(\delta_{ac}M_{bd} + \delta_{bd}M_{ac} - \delta_{ad}M_{bc} - \delta_{bc}M_{ad})\\ 
&[P_{m}, M_{ab}] =i(\delta_{ma}P_{b} - \delta_{mb}P_{a})\nonumber \\
&[P_{a}, P_{b}] = 0\nonumber \\
&[Q_{\alpha}, M_{mn}] =(\sigma_{mn})_{\alpha}^{\    \beta}Q_{\beta}\nonumber \\
&[Q_{\alpha}, P_{m}] =0 =[\bar{Q}_{\dot{\alpha}}, P_{m}]\nonumber \\
&[\bar{Q}^{\dot{\alpha}}, M_{mn}] =(\bar{\sigma}_{mn})^{\dot{\alpha}}_{\    \dot{\beta}}\bar{Q}^{\dot{\beta}}\nonumber \\
&\{Q_{\alpha}, Q_{\beta}\} = 0=\left\{\bar{Q}_{\dot{\alpha}}, \bar{Q}_{\dot{\beta}}\right\} =0\nonumber\\
&\{Q_{\alpha}, \bar{Q}_{\dot{\beta}}\} = 2(\sigma^{m})_{\alpha \dot{\beta}}(P_{m}) \nonumber 
\end{align}

Let

\begin{align}\label{10}
&A(\omega) =e^{-\frac{1}{2}\omega^{mn}\sigma_{mn}}
&B^{\dagger}(\omega) =e^{\frac{1}{2}\omega^{mn}\bar{\sigma}_{mn}}
\end{align}

be elements of $D^{(\frac{1}{2}, 0)}$ resp. $D^{(0,\frac{1}{2})} $ representations of $SU(2) \times SU(2)$ group. They obey
 $A(\omega)\sigma^{m}B^{\dagger}(\omega) =(e^{-\omega})^{m}_{\   n}\sigma^{n} $.

The commutation rules (\ref{9}) imply the following composition law for $N = 1$ Euclidean supergroup elements:

\begin{align}\label{11}
g(\xi, \bar{\xi}, a, \omega)g(\eta, \bar{\eta},b,\omega')=& g(\Lambda, \bar{\Lambda}, c, \omega'') \\
\Lambda^{\alpha} =& \xi^{\alpha} + (A^{T}(\omega))^{\alpha}_{\   \beta}\eta^{\beta}\nonumber \\ 
\bar{\Lambda}^{\dot{\alpha}} =&  \bar{\xi}^{\dot{\alpha}} + (B^{\dagger}(\omega))^{\dot{\alpha}}_{\  \dot{\beta}}\bar{\eta}^{\dot{\beta}}\nonumber \\
c^{m} =&  e(^{-\omega})^{m}_{\  n}b^{n} + a^{m} -i\xi^{\alpha}(\sigma^{m})_{\alpha \dot{\alpha}}(B^{\dagger}(\omega))^{\dot{\alpha}}_{\   \dot{\beta}}\bar{\eta}^{\dot{\beta}}\nonumber \\& -i\bar{\xi}^{\dot{\alpha}}(\bar{\sigma}^{m})_{\dot{\alpha} \alpha}(A^{T}(\omega))^{\alpha}_{\   \beta}\eta^{\beta}\nonumber\\
 e^{\frac{i}{2}\omega''_{ab}M^{ab}}=& e^{\frac{i}{2}\omega_{ab}M^{ab}}e^{\frac{i}{2}\omega'_{ab}M^{ab}}\nonumber 
\end{align}

The action of $N = 1$ Euclidean supergroup on $N = 1$ Euclidean superspace can be deduced from eq.(\ref{11})

\begin{align}\label{12}
g(\xi, \bar{\xi}, a, \omega):
(x^{m},\eta^{\alpha}, \bar{\eta}^{\dot{\alpha}})\longmapsto& (x'^{m},\eta'^{\alpha}, \bar{\eta'}^{\dot{\alpha}})\\
\eta'^{\alpha} =& \xi^{\alpha} + (A^{T}(\omega))^{\alpha}_{\   \beta}\eta^{\beta}\nonumber \\ 
\bar{\eta'}^{\dot{\alpha}} =& \bar{\xi}^{\dot{\alpha}} + (B^{\dagger}(\omega))^{\dot{\alpha}}_{\  \dot{\beta}}\bar{\eta}^{\dot{\beta}}\nonumber \\
x'^{m} =& (e^{-\omega})^{m}_{\  n}x^{n} + a^{m} -i\xi^{\alpha}(\sigma^{m})_{\alpha \dot{\alpha}}(B^{\dagger}(\omega ))^{\dot{\alpha}}_{\   \dot{\beta}}\bar{\eta}^{\dot{\beta}}\nonumber \\& -i\bar{\xi}^{\dot{\alpha}}(\bar{\sigma}^{m})_{\dot{\alpha} \alpha}(A^{T}(\omega))^{\alpha}_{\   \beta}\eta^{\beta}\nonumber
\end{align}

Now, replacing commuting coordinates $x^{m}$ and $\eta^{\alpha}$, $\bar{\eta}^{\dot{\alpha}}$ by (non)commuting quantities $\hat{x}^{m}$, $\hat{\eta}^{\alpha}$ and $ \hat{\bar{\eta}}^{\dot{\alpha}}$ satisfying some consistent commutation rules one obtains $N = 1 $ deformed Euclidean 
superspace. In general, these commutation rules are no longer covariant under the action of usual $N = 1$ Euclidean supergroup. So, the standard
Euclidean transformations have to be modified somehow if one wants the relations defining superspace deformation to be preserved under the action of these transformations. It appears that for a wide class of $N = 1$ Euclidean superspace deformations (including most of deformations considered in 
literature  \cite{a36},\cite{a55}, \cite{a56}, \cite{a57}, \cite{a58}, \cite{a59}) this can be done in relatively simple way, starting with the following intuitive and natural from a "physical" point of view assumptions: 
\begin{enumerate}[a)]
\item  In order to get generalized transformations one replace commutative parameters $(\alpha )$ by noncommutative ones $(\hat \alpha )$

\begin{align}\label{13} 
(\alpha) \equiv (\xi, \bar{\xi}, a, \omega) \longmapsto (\hat{\alpha}) \equiv (\hat{\xi}, \hat{\bar{\xi}}, \hat{a}, \hat{\omega}),  
\end{align}

In more formal approach, supergroup parameters are regarded as functions of supergroup elements and passing from commutative parameters to
(non)commmutative ones is interpreted as passing from commutative superalgebra of functions on supergroup to, in general, noncommutative algebra. \\

\item These noncommuting transformation parameters commute with deformed superspace coordinates.\\

\item  Generalized transformations act on deformed superspace as the usual ones on undeformed space \\

\begin{align}\label{14}
g(\hat{\xi,} \hat{\bar{\xi}}, \hat{a}, \hat{\omega}): &(\hat{x}^{m},\hat{\eta}^{\alpha}, \hat{\bar{\eta}}^{\dot{\alpha}})
\longmapsto (\hat{x}'^{m},\hat{\eta}'^{\alpha}, \hat{\bar{\eta}}'^{\dot{\alpha}})\\
\hat{\eta'}^{\alpha} =& \hat{\xi}^{\alpha} + (A^{T}(\hat{\omega}))^{\alpha}_{\   \beta} \hat{\eta}^{\beta}\nonumber\\ 
\hat{\bar{\eta}'}^{\dot{\alpha}} =& \hat{\bar{\xi}}^{\dot{\alpha}} + (B^{\dagger}(\hat{\omega}))^{\dot{\alpha}}_{\  \dot{\beta}}
\hat{\bar{\eta}}^{\dot{\beta}}\nonumber \\
\hat{x'}^{m} =& (e^{-\hat{\omega}})^{m}_{\  n}\hat{x}^{n} + \hat{a}^{m} -i\hat{\xi}^{\alpha}(\sigma^{m})_{\alpha \dot{\alpha}}
(B^{\dagger}(\hat{\omega}))^{\dot{\alpha}}_{\   \dot{\beta}}
\hat{\bar{\eta}}^{\dot{\beta}} \nonumber\\& -i\hat{\bar{\xi}}^{\dot{\alpha}}
(\sigma^{m})_{\alpha \dot{\alpha}}(A^{T}(\hat{\omega}))^{\alpha}_{\   \beta}\hat{\eta}^{\beta},\nonumber
\end{align}

To avoid the ordering problem one puts

\begin{align}\label{15}
[\hat{\xi}, B^{\dagger}(\hat{\omega})]=[\hat{\bar{\xi}}, A^{T}(\hat{\omega})]=0.
\end{align} \\

\item Commutation rules defining superspace deformation should be covariant under action of these generalized transformations. \\
\end{enumerate}
Assumptions (a-c) enable one to find the relations between deformed space-time coordinates $\hat{x}^{m}$, $\hat{\eta}^{\alpha}$,
 $\hat{\bar {\eta}}^{\dot \alpha}$ commutators and commutators of transformation parameters. Both, coordinates and parameters
commutation rules, remain apriori unspecified.
 
 These relations can be divided into three groups. The first one includes: commutators of grassmannian space-time 
coordinates $\hat{\eta}^{\alpha}$, $\hat{\bar {\eta}}^{\dot \alpha}$, commmutators of grassmannian transformation parameters 
$\hat{\xi }$, $\hat\bar {\xi }$, commmutators of grassmannian parameters with A, B matrices and commutator of A and B matrices.\\
There are no commmutators of space-time variables and commutators of parameters (except ones mentioned above ) in this group.\\ 
The second group of relations contains, in addition, commutators of space-time coordinates (except commutator between two bosonic elements
$\hat{x}^{m}$) and commmutators of transformation parameters (except the ones between parameters $\hat{a}^{m}$ ).\\
The commutators $[\hat{x}^{m},\;\hat{x}^{n}]$ and $[\hat{a}^{m},\;\hat{a}^{n}]$ appear in the third group of relations.\\
Such structure means that one can look for consistent commutation rules defining deformed superspaces and relevant commutation rules 
determining generalized symmetry transformations by successive analysis of these groups starting with the first one being described by the 
following equations: 
 
\begin{align}\label{16}
\{\hat{\eta}'^{\alpha}, \hat{\eta}'^{\beta}\}=&[(A^{T})^{\alpha}_{\   \gamma},(A^{T})^{\beta}_{\   \delta}]\hat{\eta}^{\gamma}\hat{\eta}^{\delta} +(A^{T})^{\beta}_{\  \delta}(A^{T})^{\alpha}_{\  \gamma}\{\hat{\eta}^{\gamma}, \hat{\eta}^{\delta}\}\\& + \hat{\eta}^{\gamma}([(A^{T})^{\alpha}_{\   \gamma},\hat{\xi}^{\beta}]+[(A^{T})^{\beta}_{\   \gamma},\hat{\xi}^{\alpha}]) + \{\hat{\xi}^{\alpha}, \hat{\xi}^{\beta}\} \nonumber  \\
\{{\hat{\bar{\eta}}}'^{\dot{\alpha}}, \hat{\bar{\eta}}'^{\dot{\beta}}\}=&[(B^{\dagger})^{\dot{\alpha}}_{\   \dot{\gamma}},
(B^{\dagger})^{\dot{\beta}}_{\   \dot{\delta}}]\hat{\bar{\eta}}^{\dot{\gamma}}\hat{\bar{\eta}}^{\dot{\delta}} 
 +(B^{\dagger})^{\dot{\beta}}_{\  \dot{\delta}}
(B^{\dagger})^{\dot{\alpha}}_{\  \dot{\gamma}}\{\hat{\bar{\eta}}^{\dot{\gamma}},
\hat{\bar{\eta}}^{\dot{\delta}}\}\nonumber\\& + \hat{\bar{\eta}}^{\dot{\gamma}}
([(B^{\dagger})^{\dot{\alpha}}_{\   \dot{\gamma}},\hat{\bar{\xi}}^{\dot{\beta}}]+[(B^{\dagger})^{\dot{\beta}}_{\   \dot{\gamma}},
\hat{\bar{\xi}}^{\dot{\alpha}}]) + \{\hat{\bar{\xi}}^{\dot{\alpha}}, \hat{\bar{\xi}}^{\dot{\beta}}\} \nonumber \\
\{\hat{\eta}'^{\alpha}, \hat{\bar{\eta}}'^{\dot{\beta}}\}=&[(A^{T})^{\alpha}_{\   \gamma},
(B^{\dagger})^{\dot{\beta}}_{\   \dot{\delta}}]\hat{\eta}^{\gamma}\hat{\bar{\eta}}^{\dot{\delta}} + (B^{\dagger})^{\dot{\beta}}_{\  \dot{\delta}}(A^{T})^{\alpha}_{\  \gamma}\{\hat{\eta}^{\gamma}, 
\hat{\bar{\eta}}^{\dot{\delta}}\}\nonumber \\& + \{\hat{\xi}^{\alpha}, \hat{\bar{\xi}}^{\dot{\beta}}\}.\nonumber
\end{align} 

Now, covariant (in view of assumption d) commutation rules of grassmannian space-time 
coordinates should be defined in such a way that eqs.  (\ref{16}) are satisfied for all space-time variables $\hat{x}^{m}$, $\hat{\eta}^{\alpha}$, $\hat{\bar{\eta}}^{\dot{\alpha}}$.\\
Consider the simplest deformation of the grassmannian sector:

\begin{eqnarray}\label{17}
\{\hat{\eta}^{\alpha}, \hat{\eta}^{\beta}\}= C^{\alpha \beta} =  \{\hat{\eta}'^{\alpha}, \hat{\eta}'^{\beta}\} \\
\{\hat{\eta}^{\alpha},\hat{\bar{\eta}}^{{\dot{\beta}}}\}= E^{\alpha \dot{\beta}} = \{\hat{\eta}'^{\alpha},\hat{\bar{\eta}}'^{{\dot{\beta}}}\}
\nonumber \\ 
\{\hat{\bar{\eta}}^{\dot{\alpha}},\hat{\bar{\eta}}^{\dot{\beta}}\}=D^{\dot{\alpha} \dot{\beta}} = \{\hat{\eta}'^{\alpha},\hat{\bar{\eta}}'^{{\dot{\beta}}}\}\nonumber
\end{eqnarray}

with $ C^{\alpha \beta} $, $ E^{\alpha \dot{\beta}}\nonumber  $ and $ D^{\dot{\alpha} \dot{\beta}} $ being the even elements of a 
Grassmann algebra, in particular, c-numbers. \\
Then, the sufficient conditions for eqs.(\ref{16}) to hold read 

\begin{align}\label{18}
&[(A^{T})^{\alpha}_{\   \gamma},(A^{T})^{\beta}_{\   \delta}]=0\nonumber\\
&[(B^{\dagger})^{\dot{\alpha}}_{\   \dot{\gamma}},(B^{\dagger})^{\dot{\beta}}_{\   \dot{\delta}}]=0\nonumber \\
&[(B^{\dagger})^{\dot{\alpha}}_{\   \dot{\gamma}},(A^{T})^{\beta}_{\   \delta}]=0\nonumber \\
&[\hat{\xi}^{\alpha},(A^{T})^{\beta}_{\   \delta}]=0\nonumber\\
&[(B^{\dagger})^{\dot{\alpha}}_{\   \dot{\gamma}},\hat{\bar{\xi}}^{\dot{\beta}}]=0\nonumber \\
&\{\hat{\xi}^{\alpha}, \hat{\xi}^{\beta} \}= C^{\alpha \beta}_{-}\nonumber\\
&\{\hat{\bar{\xi}}^{\dot{\alpha}}, \hat{\bar{\xi}}^{\dot{\beta}} \}= D^{\dot{\alpha} \dot{\beta}}_{-}\nonumber\\
&\{\hat{\xi}^{\alpha}, \hat{\bar{\xi}}^{\dot{\beta}} \}= E^{\alpha \beta}_{-}\nonumber \\
\end{align}

where $C^{\alpha \beta}_{-}$, etc. are defined as follows

\begin{align}\label{19}
&C^{\alpha \beta}_{\pm}=[\delta^{\alpha}_{\gamma}\delta^{\beta}_{\delta}\pm(A^{T})^{\alpha}_{\  \gamma}(A^{T})^{\beta}_{\   \delta}]C^{\gamma \delta}\\
&D^{\dot{\alpha} \dot{\beta}}_{\pm}=[\delta^{\dot{\alpha}}_{\dot{\gamma}}\delta^{\dot{\beta}}_{\dot{\delta}}
\pm(B^{\dagger})^{\dot{\alpha}}_{\ \dot{\gamma}}(B^{\dagger})^{\dot{\beta}}_{\   \dot{\delta}}]D^{\dot{\gamma} \dot{\delta}}\nonumber \\
&E^{\alpha \dot{\beta}}_{\pm}=[\delta^{\alpha}_{\gamma}\delta^{\dot{\beta}}_{\dot{\delta}}
\pm(A^{T})^{\beta}_{\   \gamma}(B^{\dagger})^{\dot{\beta}}_{\  \dot{\delta}}]E^{\gamma \dot{\delta}}\nonumber 
\end{align}

Eqs. (\ref{19}) define also $C^{\alpha \beta}_{+}$, etc. which will be used below.

Taking into account commutation rules satisfied by $A^{T}$ and $B^{+}$ (see eqs.(\ref{18})) allows one to write out second group of equations
relating commutators of spacetime variables with commutators of parameters 

\begin{align}\label{20}
[\hat{x}'^{m}, \hat{\eta}'^{\alpha}] = &(e^{-\hat{\omega}})^{m}_{\   n}(A^{T})^{\alpha}_{\   \gamma}[\hat{x}^{n},\hat{\eta}^{\gamma}] +[\hat{a}^{m}, \hat{\xi}^{\alpha}] +[\hat{a}^{m}, (A^{T})^{\alpha}_{\  \beta}]\hat{\eta}^{\beta}\\& + i\sigma^{m}_{\delta \dot{\delta}}
(B^{\dagger})^{\dot{\delta}}_{\   \dot{\beta}}
(\{\hat{\xi}^{\delta}, \hat{\xi}^{\alpha}\}
\hat{\bar{\eta}}^{\dot{\beta}} - (A^{T})^{\alpha}_{\   \gamma}\hat{\xi}^{\delta}\{\hat{\bar{\eta}}^{\dot{\beta}},\hat{\eta}^{\gamma}\}) \nonumber \\
&+i\sigma^{m}_{\delta \dot{\delta}}(A^{T})^{\delta}_{\   \beta}(\{\hat{\bar{\xi}}^{\dot{\delta}}, \hat{\xi}^{\alpha}\}
\hat{\eta}^{\beta} - (A^{T})^{\alpha}_{\   \gamma}\hat{\bar{\xi}}^{\dot{\delta}}\{\hat{\eta}^{\dot{\beta}},\hat{\eta}^{\gamma}\}) \nonumber\\
[\hat{x}'^{m}, \hat{\bar{\eta}}'^{\dot{\alpha}}] = &(e^{-\hat{\omega}})^{m}_{\   n}
(B^{\dagger})^{\dot{\alpha}}_{\   \dot{\gamma}}[\hat{x}^{n},\bar{\eta}^{\dot{\gamma}}]
 +[\hat{a}^{m}, \hat{\bar{\xi}}^{\dot{\alpha}}] +[\hat{a}^{m}, (B^{\dagger})^{\dot{\alpha}}_{\  \dot{\beta}}]\hat{\bar{\eta}}^{\dot{\beta}}\nonumber \\& + i\sigma^{m}_{\delta \dot{\delta}}(A^{T})^{\delta}_{\   \beta}(\{\bar{\xi}^{\dot{\delta}}, \hat{\bar{\xi}}^{\dot{\alpha}}\}
\hat{\eta}^{\beta} - (B^{\dagger})^{\dot{\alpha}}_{\   \dot{\gamma}}\hat{\bar{\xi}}^{\dot{\delta}}
\{\hat{\eta}^{\beta},\hat{\bar{\eta}}^{\dot{\gamma}}\}) \nonumber \\
&+i\sigma^{m}_{\delta \dot{\delta}}
(B^{\dagger})^{\dot{\delta}}_{\   \dot{\beta}}(\{\hat{\xi}^{\delta}, {\hat{\bar{\xi}}}^{\dot{\alpha}}\}
\hat{\eta}^{\beta} - (B^{\dagger})^{\dot{\alpha}}_{\   \dot{\gamma}}\hat{\xi}^{\delta}\{\hat{\eta}^{\dot{\beta}},
\hat{\bar{\eta}}^{\dot{\gamma}}\})\nonumber  
\end{align}

Again, these equations should hold for all spacetime variables. This imposes some constraints on covariant commutation rules of $\hat{\eta}^{\alpha}$, $\hat{x}^{m}$ and $\hat{\bar{\eta}}^{\dot{\alpha}}$, $\hat{x}^{m}$ coordinates. In fact, it is not difficult to see that these rules should be at least linear in $\hat{\eta}^{\alpha}$ and $\hat{\bar{\eta}}^{\dot{\alpha}}$ variables. Assuming that they are exactly linear one can write

\begin{align}\label{21}
&[\hat{x}^{m}, \hat{\eta}^{\alpha}]=i\Pi^{m\alpha}_{\   \beta}\hat{\eta}^{\beta} + i\Pi^{m\alpha}_{\   \dot{\beta}}\hat{\bar{\eta}}^{\dot{\beta}}+ i\Pi^{m \alpha}\\
&[\hat{x}^{m}, \hat{\bar{\eta}}^{\dot{\alpha}}]=i\Delta^{m \dot{\alpha}}_{\   \beta}\hat{\eta}^{\beta} + i\Delta^{m \dot{\alpha}}_{\   \dot{\beta}}\hat{\bar{\eta}}^{\dot{\beta}}+ i\Delta^{m \dot{\alpha}}\nonumber
\end{align}

where  $\Pi^{m\alpha}_{\   \beta}$, $\Pi^{m\alpha}_{\   \dot{\beta}}$, $\Pi^{m \alpha}$, $\Delta^{m \dot{\alpha}}_{\   \beta}$, $\Delta^{m \dot{\alpha}}_{\   \dot{\beta}}$, $\Delta^{m \dot{\alpha}}$ are some constants of well-defined parity.\\ 
Now, taking into account eqs. (\ref{18}), one can provide the sufficient conditions for eqs.(\ref{20}) to be satisfied by all space-time variables. 
They have the form of the following relations between the constants $\Pi$, $\Delta$ and $E$, $D$ , $C$

\begin{align}\label{22}  
&\Pi^{m\alpha}_{\   \beta}=E^{\alpha \dot{\beta}}(\sigma^{m})_{\beta \dot{\beta}}\\
&\Pi^{m\alpha}_{\   \dot{\beta}}=C^{\alpha \beta}(\sigma^{m})_{\beta \dot{\beta}}\nonumber \\
&\Delta^{m \dot{\alpha}}_{\   \beta}=D^{{\dot{\alpha}} \dot{\beta}}(\sigma^{m})_{\beta \dot{\beta}}\nonumber \\
&\Delta^{m \dot{\alpha}}_{\   \dot{\beta}}=E^{ \beta \dot{\alpha}}(\sigma^{m})_{\beta \dot{\beta}}\nonumber
\end{align}

and the following commutation rules for transformation parameters  

\begin{align}\label{23} 
[(A^{T})^{\alpha}_{\   \beta}, \hat{a}^{m}] =& 0\nonumber\\
[(B^{\dagger})^{\dot{\alpha}}_{\   \dot{\beta}}, \hat{a}^{m}] =& 0\nonumber \\
[\hat{a}^{m}, \hat{\xi}^{\alpha}]=&iE_{+}^{\alpha \dot{\rho}}(\sigma^{m})_{\rho \dot{\rho}}\hat{\xi}^{\rho} + 
iC_{+}^{\alpha \rho}(\sigma^{m})_{\rho \dot{\rho}}\hat{\bar{\xi}}^{\dot{\rho}}+ i\Pi^{m \alpha}_{-}\nonumber \\
[\hat{a}^{m}, \hat{\bar{\xi}}^{\dot{\alpha}}]=&iE^{ \rho \dot{\alpha}}_{+}(\sigma^{m})_{\rho \dot{\rho}}\hat{\bar{\xi}}^{\dot{\rho}} + 
iD^{\dot{\alpha}\dot{\rho}}_{+}(\sigma^{m})_{\rho \dot{\rho}}\hat{\xi}^{\rho}+ i\Delta^{m \dot{\alpha}}_{-}\nonumber \\
\end{align}

In the eqs. (\ref{23}) $\Delta^{m \dot{\alpha}}_{-}$ and $\Pi^{m \alpha}_{-}$ are defined by

\begin{align}\label{24} 
&\Delta^{m \dot{\alpha}}_{\pm}=(\delta^{m}_{p}\delta^{\dot{\alpha}}_{\dot{\beta}}
 \pm(e^{-\hat{\omega}})^{m}_{\   p}(B^{\dagger})^{\dot{\alpha}}_{\   \dot{\beta}})\Delta^{p \dot{\beta}}\nonumber\\
&\Pi^{m \alpha}_{\pm}=(\delta^{m}_{p}\delta^{\alpha}_{\beta} \pm(e^{-\hat{\omega}})^{m}_{\   p}(A^{T})^{\alpha}_{\    \beta})\Pi^{p \beta}
\end{align}
 
It follows from eqs. (\ref{15}), (\ref{18}) and (\ref{23})  that the commutativity of $A^{T}$ and $B^{+}$ with all parameters of transformations
(\ref{14}) is consistent with the basic assumptions (a-d).
Using that, last group of equations relating commutators of $\hat{\xi}$ coordinates and commutators of $\hat {a}^{m}$ parameters can be written in 
the following form

\begin{align}\label{25}
[\hat{x}'^{m},\hat{x}'^{n}]&=(e^{-\hat{\omega}})^{m}_{\  p}(e^{-\hat{\omega}})^{n}_{\  q}[\hat{x}^{p}, \hat{x}^{q}] +[\hat{a}^{m}, \hat{a}^{n}]    \\& -i((e^{-\hat{\omega}})^{m}_{\  p}(\sigma^{n})_{\alpha \dot{\alpha}} -(e^{-\hat{\omega}})^{n}_{\  p}(B^{\dagger})^{\dot{\alpha}}_{\  \dot{\beta}}\hat{\xi}^{\alpha}[\hat{x}^{p}, \hat{\bar{\eta}}^{\dot{\beta}}] + (A^{T})^{\alpha}_{\  \beta}\hat{\bar{\xi}}^{\dot{\alpha}}[\hat{x}^{p}, \hat{\eta}^{\beta}]\nonumber \\& 
- i((\sigma^{n})_{\alpha \dot{\alpha}}[\hat{a}^{m}, \hat{\xi}^{\alpha}]-(\sigma^{m})_{\alpha \dot{\alpha}}[\hat{a}^{n}, \hat{\xi}^{\alpha}])(B^{\dagger})^{\dot{\alpha}}_{\   \dot{\beta}}\hat{\bar{\eta}}^{\dot{\beta}}\nonumber \\&
 - i((\sigma^{n})_{\alpha \dot{\alpha}}[\hat{a}^{m}, \hat{\bar{\xi}}^{\dot{\alpha}}]-(\sigma^{m})_{\alpha \dot{\alpha}}[\hat{a}^{n}, \hat{\bar{\xi}}^{\dot{\alpha}}])(A^{T})^{\alpha}_{\   \beta}\hat{\eta}^{\beta}\nonumber \\&
 + \frac{1}{2}(\sigma^{m})_{\gamma \dot{\gamma}}(\sigma^{n})_{\delta \dot{\delta}}(B^{\dagger})^{\dot{\gamma}}_{\   \dot{\alpha}}(B^{\dagger})^{\dot{\delta}}_{\   \dot{\beta}}([\hat{\xi}^{\gamma},\hat{\xi}^{\delta}]\{\hat{\bar{\eta}}^{\dot{\alpha}},\hat{\bar{\eta}}^{\dot{\beta}}\}-\{\hat{\xi}^{\gamma},\hat{\xi}^{\delta}\}[\hat{\bar{\eta}}^{\dot{\alpha}},\hat{\bar{\eta}}^{\dot{\beta}}])\nonumber \\& +
\frac{1}{2}(\sigma^{m})_{\gamma \dot{\gamma}}(\sigma^{n})_{\delta \dot{\delta}}(A^{T})^{\gamma}_{\  \alpha}(A^{T})^{\delta}_{\   \beta}([\hat{\bar{\xi}}^{\dot{\gamma}},\hat{\bar{\xi}}^{\dot{\delta}}]\{\hat{\eta}^{\alpha},\hat{\eta}^{\beta}\}-\{\hat{\bar{\xi}}^{\dot{\gamma}},\hat{\bar{\xi}}^{\dot{\delta}}[\hat{\eta}^{\alpha},\hat{\bar{\eta}}^{\dot{\beta}}])\nonumber \\& +
\frac{1}{2}((\sigma^{m})_{\gamma \dot{\gamma}}(\sigma^{n})_{\delta \dot{\delta}} - (\sigma^{n})_{\gamma \dot{\gamma}}(\sigma^{m})_{\delta \dot{\delta}} (B^{\dagger})^{\dot{\gamma}}_{\  \dot{\alpha}}(A^{T})^{\delta}_{\   \beta})([\hat{\xi}^{\gamma},\hat{\bar{\xi}}^{\dot{\delta}}]\{\hat{\bar{\eta}}^{\dot{\alpha}},\hat{\eta}^{\beta}\}\nonumber \\&-\{\hat{\xi}^{\gamma},\hat{\bar{\xi}}^{\dot{\delta}}\}[\hat{\eta}^{\beta},\hat{\bar{\eta}}^{\dot{\alpha}}]) \nonumber 
\end{align}

All commutation relations in eqs.(\ref{25}), except $[\hat{x}^{m},\hat{x}^{n}]$ and $[\hat{a}^{m},\hat{a}^{n}]$ commutators, have been already determined (see eqs. (\ref{17}), (\ref{18}), (\ref{19}), (\ref{21}), (\ref{22}) and (\ref{23})).
Requiring eqs. (\ref{25}) to be satisfied by all space-time coordinates $\hat{\eta}^{\alpha}$, $\hat{\bar{\eta}}^{\dot{\alpha}}$, $\hat{x}^{m}$ implies that the commutator $[\hat{x}^{m},\hat{x}^{n}]$ should be at least quadratic in $\hat{\eta}^{\alpha}$, $\hat{\bar{\eta}}^{\dot{\alpha}}$ variables. We assume the following general form of the commutator   

\begin{align}\label{26}
[\hat{x}^{m},\hat{x}^{n}]=\Gamma^{mn}_{  \gamma \delta}[\hat{\eta}^{\gamma},\hat{\eta}^{\delta}] + \Gamma^{mn}_{  \dot{\gamma} \dot{\delta}}[\hat{\bar{\eta}}^{\dot{\gamma}},\hat{\bar{\eta}}^{\dot{\delta}}] +  \Gamma^{mn}_{  \delta \dot{\gamma}}[\hat{\eta}^{\delta},\hat{\bar{\eta}}^{\dot{\gamma}}] + \Gamma^{mn}_{   \delta}\hat{\eta}^{\delta}  + \Gamma^{mn}_{   \dot{\delta}}\hat{\bar{\eta}}^{\dot{\delta}} + i\Theta^{mn},
\end{align}

where  $\Gamma^{mn}_{  \gamma \delta}$, $ \Gamma^{mn}_{  \dot{\gamma} \dot{\delta}}$, $\Gamma^{mn}_{  \delta \dot{\gamma}}$, $\Gamma^{mn}_{  \delta}$, $\Gamma^{mn}_{  \dot{\delta}}$ and $\Theta^{mn}$ are some constants of well defined parity. Inserting eq.(\ref{26}) into eqs. (\ref{24}) allows one to express  
 $\Gamma^{mn}_{  \gamma \delta}$, $ \Gamma^{mn}_{  \dot{\gamma} \dot{\delta}}$, $\Gamma^{mn}_{\delta \dot{\gamma}}$, 
$\Gamma^{mn}_{  \delta}$, $\Gamma^{mn}_{  \dot{\delta}}$ in terms of $C^{\alpha \beta }$, $D^{\dot \alpha \dot \beta }$ and $E^{\alpha \dot \beta }$; 

\begin{align}\label{27}
&\Gamma^{mn}_{  \gamma \delta} =\frac{1}{2} D^{\dot{\alpha} \dot{\beta}} (\sigma^{m})_{\gamma \dot{\alpha}}(\sigma^{n})_{\delta \dot{\beta}}\\
&\Gamma^{mn}_{  \dot{\gamma} \dot{\delta}}=\frac{1}{2} C^{\alpha \beta} (\sigma^{m})_{\alpha \dot{\gamma}}(\sigma^{n})_{\beta \dot{\delta}}\nonumber\\
&\Gamma^{mn}_{  \delta \dot{\gamma}}=\frac{1}{2} E^{\alpha \dot{\beta}} ((\sigma^{m})_{\delta \dot{\beta}}(\sigma^{n})_{\alpha \dot{\gamma}} - (\sigma^{n})_{\delta \dot{\beta}}(\sigma^{m})_{\alpha \dot{\gamma}})\nonumber \\
&\Gamma^{mn}_{  \dot{\delta}}=\Pi^{m \alpha}(\sigma^{n})_{\alpha \dot{\delta}} - \Pi^{n \alpha}(\sigma^{m})_{\alpha \dot{\delta}}\nonumber\\
&\Gamma^{mn}_{  \delta}=\Delta^{m \dot{\alpha}}(\sigma^{n})_{\delta \dot{\alpha}} - \Delta^{n \dot{\alpha}}(\sigma^{m})_{\delta \dot{\alpha}}\nonumber
\end{align}

Moreover, one also finds  the form of $[\hat{a}^{m},\hat{a}^{n}]$ commutators

\begin{align}\label{28}
 [\hat{a}^{m},\hat{a}^{n}] =& 
\frac{1}{2} D_{-}^{\dot{\alpha} \dot{\beta}} 
(\sigma^{m})_{\gamma \dot{\alpha}}(\sigma^{n})_{\delta \dot{\beta}}[\hat{\xi}^{\gamma}, \hat{\xi}^{\delta}] \\
&+\frac{1}{2} C_{-}^{\alpha \beta} (\sigma^{m})_{\alpha \dot{\gamma}}
(\sigma^{n})_{\beta \dot{\delta}}[\hat{\bar{\xi}}^{\dot{\gamma}}, \hat{\bar{\xi}}^{\dot{\delta}}]\nonumber \\
& + \frac{1}{2} E_{-}^{\alpha \dot{\beta}} ((\sigma^{m})_{\gamma \dot{\beta}}(\sigma^{n})_{\alpha \dot{\delta}} - (\sigma^{n})_{\gamma \dot{\beta}}(\sigma^{m})_{\alpha \dot{\delta}})[\hat{\xi}^{\gamma}, \hat{\bar{\xi}}^{\dot{\delta}}] \nonumber\\
&+ (\Delta_{+}^{m \dot{\alpha}}(\sigma^{n})_{\sigma \dot{\alpha}} - \Delta_{+}^{n \dot{\alpha}}(\sigma^{m})_{\sigma \dot{\alpha}})\hat{\xi}^{\sigma}\nonumber \\
& + (\Pi_{+}^{m \alpha}(\sigma^{n})_{\alpha \dot{\sigma}} - \Pi^{n \alpha}_{+}(\sigma^{m})_{\alpha \dot{\sigma}})\hat{\bar{\xi}}^{\dot{\sigma}} + i\Theta^{mn}_{-}\nonumber
\end{align}

where $\Theta^{mn}_{-}$ is given by

\begin{align}\label{29}
\Theta^{mn}_{-} = (\delta^{m}_{p}\delta^{n}_{q}-(e^{-\hat{\omega}})^{m}_{\  p}(e^{-\hat{\omega}})^{n}_{\  q})\Theta^{pq}
\end{align}

So, finally, we conclude that the commutation rules defining deformed $N=1$ Euclidean superspace, given by eqs. (\ref{17}), 
(\ref{21}), (\ref{22}), (\ref{26}), (\ref{27}), are covariant 
under the global space-time transformations (\ref{14}), with the parameters $(\hat{\xi}, \hat{\bar{\xi}}, \hat{a}, \hat{\omega})$ satisfying 
the commutation rules described by eqs. (\ref{15}), (\ref{18}), (\ref{19}), (\ref{23}), (\ref{24}), (\ref{28}) and (\ref{29}).\\
It is interesting to note that coordinates commutators and those of parameters are similar. Also we see that the choice of grassmannian
superspace sector deformations determines to much extent the remaining commutation rules, both for coordinates and parameters.\\
The discussion  of superspace deformations and their symmetries would not be complete without considering (super)-Jacobi identities and 
superalgebraic structure of these symmetries.\\
It appears that, in general, without further constraints on C, D, E, $\Pi $ and $\Delta $ matrices the commutation rules for coordinates 
as well as for parameters, found above, are not consistent with Jacobi identities. In fact, the ones involving: two grassmannian and one bosonic, 
two bosonic and one grassmannian and finally, three bosonic space-time coordinates are not satisfied automatically if C, D, E matrices are arbitrary
(for details concerning Jacobi identities see Appendix). Imposing Jacobi identities leads to a very complicated system of equations on the elements of these matrices. The simplest non-trivial solutions to these equations read
 
\begin{enumerate}[a)]
\item
\begin{align}\label{30}
&C^{\alpha \beta} \neq 0,& &\Pi^{m \beta} \neq 0,&  &\Theta^{mn} \neq 0\\
&D^{\dot{\alpha} \dot{\beta}}=0,&  &E^{\beta \dot{\alpha}}=0,& &\Delta^{m \dot{\beta}}=0\nonumber
\end{align}

with $C^{\alpha \beta}$, $\Theta^{mn}$ being arbitrary c-numbers or even grassmannian constants and $\Pi^{m \beta}$ - arbitrary odd grassmannian ones and 

\item
\begin{align}\label{31}
&C^{\alpha \beta} = 0,&        &\Pi^{m \beta}=0,& &\Theta^{mn} \neq 0\\
&D^{\dot{\alpha} \dot{\beta}}\neq 0,&  &E^{\beta \dot{\alpha}}=0,& &\Delta^{m \dot{\beta}}\neq 0,\nonumber
\end{align}
\end{enumerate}

with
$D^{\dot{\alpha} \dot{\beta}}$, $\Theta^{mn}$ being arbitrary even grassmannian constants (in particular, c-numbers)
and $\Delta^{m \dot{\beta}}$ - arbitrary odd grassmannian ones.\\
The solutions (\ref{30}) (resp.(\ref{31})) describe the deformation of undotted (resp. dotted) grassmannian sectors.

Deformation with $C^{\alpha \beta}\neq 0$ defining the so called $N = \frac{1}{2}$ supersymmetry and its extensions 
have been analyzed intensively in literature \cite{a55}, \cite{a56}, \cite{a57}, \cite{a58} ( in particular, also by Seiberg in ref.\cite{a45}).\\
It can be shown that if the elements of C, D, E matrices are c-numbers the solutions a/b given by eqs. (\ref{30})/(\ref{31}) 
are the only non-trivial ones which do not impose any further constraints on constants $C$, $\Pi/$D$,\Delta$. (for the proof see Appendix ). 
On the other hand, allowing the matrix elements to be even elements of some Grassmann algebra enables one to construct non-trivial matrices C, D, E defining commutation rules consistent with Jacobi identities.\\
The analysis of Jacobi identities for parameters of transformations can be performed in a similar way as in the case of coordinates because the constants
$C^{\alpha \beta}_{\pm}$, $D^{\dot{\alpha} \dot{\beta}}_{\pm}$, $E^{\beta \dot{\alpha} }_{\pm}$, $\Pi^{m \beta}_{\pm}$, $\Delta^{m \dot{\beta}}_{\pm}$
appearing in commutators of parameters differ from those entering the coordinates commutators by factors depending only on matrices
$A^{T}$, $B^{\dagger}$ and $e^{-\hat{\omega}}$ which commute with all parameters. In particular, the constants given by eqs.(\ref{30}) or((\ref{31}) define commutation rules for parameters which are consistent with Jacobi identities.        

Finally, let us consider the coalgebraic structure of symmetry transformations. In superspace $SC(\mathcal{E})$ of functions on Euclidean supergroup the relevant mappings \\

\begin{align}\label{32}
&\mbox{co-product \ \ }  \Delta: SC(\mathcal{E}) \longmapsto \   SC(\mathcal{E}) \otimes  SC(\mathcal{E})\\
&\mbox{co-unit \ \ } \varepsilon:SC(\mathcal{E}) \longmapsto\  \ \mathbb{C}\nonumber \\
&\mbox{antipode\ \ } \ S: SC(\mathcal{E}) \longmapsto\  \ SC(\mathcal{E})\nonumber
\end{align}
can be defined as follows\\

\begin{align}\label{33}
&\Delta(e^{-\omega})^{p}_{\  q} = (e^{-\omega})^{p}_{\  m} \otimes (e^{-\omega})^{m}_{\  q}\\
&\Delta(\xi^{\gamma}) = \xi^{\gamma} \otimes I + (A^T)^{\gamma}_{\     \delta} \otimes \xi^{\delta}\nonumber \\
&\Delta(\bar{\xi}^{\dot{\gamma}}) = \bar{\xi}^{\dot{\gamma}} \otimes I + (B^{\dagger})^{\dot{\gamma}}_{\     \dot{\delta}} \otimes \bar{\xi}^{\dot{\delta}}\nonumber \\
&\Delta(a^{p}) = (e^{-\omega})^{p}_{\  q} \otimes a^{q} + a^{p} \otimes I
-i\xi^{\alpha}(\sigma^{p})_{\alpha \dot{\alpha}}(B^{\dagger})^{\dot{\alpha}}_{\    \dot{\beta}} \otimes \bar{\xi}^{\dot{\beta}} - i\bar{\xi}^{\dot{\alpha}}( \bar{\sigma}^{m})_{\dot{\alpha} \alpha} (A^{T})^{\alpha}_{\  \gamma} \otimes \xi^{\gamma} \nonumber \\
&\Delta((A^{T})^{\alpha}_{\  \beta}) = (A^{T})^{\alpha}_{\  \gamma} \otimes (A^{T})^{\gamma}_{\  \beta}\nonumber \\
&\Delta((B^{\dagger})^{\dot{\alpha}}_{\  \dot{\beta}}) = (B^{\dagger})^{\dot{\alpha}}_{\  \dot{\gamma}} \otimes (B^{\dagger})^{\dot{\gamma}}_{\  \dot{\beta}}\nonumber
\end{align}

\begin{align}\label{34}
&\varepsilon((e^{-\omega})^{p}_{\  q}) = \delta^{p}_{q}\\
&\varepsilon(a^{p}) = 0 \nonumber\\
&\varepsilon((B^{\dagger})^{\dot{\alpha}}_{\   \dot{\beta}}) = \delta^{\dot{\alpha}}_{\dot{\beta}}\nonumber \\
&\varepsilon((A^{T})^{\alpha}_{\   \beta}) = \delta^{\alpha}_{\beta}\nonumber \\
&\varepsilon(\xi^{\alpha}) = 0 \nonumber \\
&\varepsilon(\bar{\xi}^{\dot{\alpha}}) = 0 \nonumber
\end{align}

\begin{align}\label{35}
&S((e^{-\omega})^{p}_{\  q}) = (e^{\omega})^{p}_{\   q} \\
&S(a^{m}) = -(e^{-\omega})^{m}_{\  n}a^{n} \nonumber\\
&S(\xi^{\alpha}) = -[(A^{T})^{-1}]^{\alpha}_{\  \beta}\xi^{\beta}\nonumber \\
&S(\bar{\xi}^{\dot{\alpha}}) = -[(B^{\dagger})^{-1}]^{\dot{\alpha}}_{\  \dot{\beta}}\bar{\xi}^{\dot{\beta}}\nonumber\\
&S((A^{T})^{\alpha}_{\   \beta}) = [(A^{T})^{-1}]^{\alpha}_{\   \beta}\nonumber\\
&S((B^{\dagger})^{\dot{\alpha}}_{\   \dot{\beta}}) = [(B^{\dagger})^{-1}]^{\dot{\alpha}}_{\   \dot{\beta}}\nonumber
\end{align}

Direct check shows that the coproduct and counit maps are graded homomorphisms of superalgebra while antipode map is its graded
anti-homomorphism i.e.

\begin{align}\label{36}
&\Delta(ab)=\Delta(a)\Delta(b)\\
&\varepsilon(ab)=\varepsilon(a)\varepsilon(b)\nonumber \\
&S(ab)=(-1)^{\left|a\right|\left|b\right|}S(b)S(a),\nonumber    
\end{align}
(here $\left|a\right|$, $\left|b\right| = 0,1$  denote the parities of a,b elements).\\
They satisfy the following identities:\\

\begin{align}\label{37}
&(I \otimes \Delta)\Delta(a) = (\Delta \otimes I)\Delta(a) \\
&(I \otimes \varepsilon)\Delta(a)=(\varepsilon \otimes I)\Delta(a) = a \nonumber \\
&m(I \otimes S)\Delta(a) = m(S \otimes I)\Delta(a)=\varepsilon(a)I\nonumber
\end{align}
where m denotes the multiplication homeomorphism on the Hopf algebra. The multiplication of elements with given parity in the 
superspace tensor product is given by equation
$(a \otimes b)(c \otimes d) =(-1)^{\left|b\right|\left|c\right|}(ac \otimes bd)$\\
This confirms that $\Delta$, S and $\varepsilon$ maps are well defined. What is more, one can verify that for the deformations
considered above which obey Jacobi identities (in particular those given by eqs. (\ref{30}),(\ref{31})) both structures, superalgebraic and supercoalgebraic ones, are consistent. This guarantees that the composition of two generalized symmetry transformations depending on two 
commuting sets of parameters is again generalized symmetry transformation with parameters satisfying the relevant commutation rules.
That, in turn, means that one deals with noncommutative Hopf superalgebras i.e. Euclidean quantum supergroups which can be regarded as supersymmetric extensions of $\theta $ - Euclidean group. This is the case for generalized transformations preserving the $N = \frac{1}{2}$ supersymmetry and its extensions given by eq.(\ref{30}). The supersymmetric generalizations of $\Theta$-Euclidean group generated by these transformations can be considered as global counterparts of appropriately twisted Euclidean superalgebra studied in \cite{a55}, \cite{a56}, 
\cite{a57}, \cite{a58}.

\subsection{N=1 deformed Minkowski Superspace }
 
N=1 deformed Minkowski superspaces and their symmetries can be introduced and analyzed in a similar way as Euclidean ones. However, in the case of Poincare supergroup, supertranslation generators $Q_{\alpha}$ and $\bar{Q}_{\dot{\alpha}}$ transform according to two fundamental representations of   
$SL(2,\mathbb{C})$ group  (universal covering of Lorentz group ) which are related by hermitian conjugation transformation. Hence, in Poincare supergroup $(Q_{\alpha})^{\dagger}$ = $\bar{Q}_{\dot{\alpha}}$. Such condition does not have to hold for supertranslation generators in Euclidean case
where two fundamental representations of $SU(2)$ $\times$ $SU(2)$ group (universal covering of 4D rotation group ) are independent.
It appears that this apparently minor difference has significant consequences making the Minkowski superspace much more deformation-resist then Euclidean one.\\
 Taking this difference into account  one can write out a generic element $g(\xi, \bar{\xi}, a, \omega)$
 of $N = 1$ Poincare supergroup in a similar way as in the Euclidean case.
To this end let $P_{m}$, resp $M_{mn}$, $ m =0,1,2,3$ be the generators of translations ( $a^{m}$ ), resp. Lorentz transformations ( $\omega^{mn}$ = $-\omega^{nm}$ ) in Minkowski space, while $ Q_{\alpha},\  \ \bar{Q}_{\dot{\alpha}} ,\   \ \alpha =1,2 ,\  \dot{\alpha} =\dot{1},\dot{2}$ -odd generators of supertranslations ($\xi^{\alpha}$, $\bar{\xi}^{\dot{\alpha}}$). Then        

\begin{align}
g(\xi, \bar{\xi}, a, \omega) &=exp\left\{i(\xi^{\alpha}Q_{\alpha} + \bar{Q}_{\dot{\alpha}}\bar{\xi}^{\dot{\alpha}})\right\}exp\{ia^{m}P_{m}\}
exp\{\frac{i}{2}\omega^{mn}M_{mn} \}
\end{align}

The commuting  $a^{m}$, $\omega^{mn}$ and anticommuting $\xi^{\alpha}$, $\bar{\xi}^{\dot{\alpha}}$ group parameters should verify the following relations 

\begin{align}\label{39}
&(\xi^{\alpha})^{\ast} = \bar{\xi}^{\dot{\alpha}}\\
&(a^{m})^{\ast} =a^{m}\nonumber \\
&(\omega^{mn})^{\ast} =\omega^{mn}\nonumber \\
&(\xi^{\alpha}\xi^{\beta})^{\ast} = \bar{\xi}^{\dot{\beta}}\bar{\xi}^{\dot{\alpha}}\nonumber \\
&(a^{m}\xi^{\alpha})^{\ast}=\bar{\xi}^{\dot{\alpha}}a^{m}\nonumber \\
&(a^{m}a^{n})^{\ast} =(a^{n}a^{m})\nonumber
\end{align}
The counterparts of definitions (\ref{7}),(\ref{8}) read ( $g_{mn} = diag(1,-1,-1,-1)$ )

\begin{align}\label{40}
&\sigma_{m} =(I,\sigma_{i})\\
&\bar{\sigma}_{m} =(I, -\sigma_{i}),\nonumber
\end{align}

\begin{align}\label{41}
\sigma_{mn} = \frac{i}{4}(\sigma_{m}\bar{\sigma}_{n}-\sigma_{n}\bar{\sigma}_{m}) \\
\bar{\sigma}_{mn} = \frac{i}{4}(\bar{\sigma}_{m}\sigma_{n}-\bar{\sigma}_{n}\sigma_{m}).\nonumber 
\end{align}

$\sigma_{mn}$, (resp. $\bar{\sigma}_{mn}$) are the generators of $D^{(\frac{1}{2},0)}$ (resp.$D^{(0,\frac{1}{2})}$) representations of 
$sl(2,\mathbb{C})$.
The generators $Q_{\alpha}$, $\bar{Q}_{\dot{\alpha}}$, $M_{mn}$, $P_{m}$ satisfy the following commutation rules\\

\begin{align}\label{42} 
&[M_{ab}, M_{cd}] = -i(g_{ac}M_{bd} + g_{bd}M_{ac} - g_{ad}M_{bc} - g_{bc}M_{ad}) \\ 
&[P_{m}, M_{ab}] =i(g_{ma}P_{b} - g_{mb}P_{a}) \nonumber \\
&[P_{a}, P_{b}] = 0\nonumber  \\
&[Q_{\alpha}, M_{mn}] =(\sigma_{mn})_{\alpha}^{\    \beta}Q_{\beta}\nonumber \\
&[\bar{Q}^{\dot{\alpha}}, M_{mn}] =(\bar{\sigma}_{mn})^{\dot{\alpha}}_{\    \dot{\beta}}\bar{Q}^{\dot{\beta}}\nonumber\\
&\{Q_{\alpha}, Q_{\beta}\} =0\nonumber \\
&\{\bar{Q}_{\dot{\alpha}}, \bar{Q}_{\dot{\beta}}\} =0\nonumber\\
&\{Q_{\alpha}, \bar{Q}_{\dot{\beta}}\} = 2(\sigma^{m})_{\alpha \dot{\beta}}P_{m}\nonumber \\ 
&[Q_{\alpha}, P_{m}] =0\nonumber \\
&[\bar{Q}_{\dot{\alpha}}, P_{m}]=0,\nonumber
\end{align}

Let

\begin{align}\label{43}
&A(\omega) =e^{-\frac{1}{2}\omega^{mn}\sigma_{mn}}
&A^{\dagger}(\omega) =e^{\frac{1}{2}\omega^{mn}\bar{\sigma}_{mn}}
\end{align}
be the elements of $D^{(\frac{1}{2}, 0)}$ resp. $D^{(0,\frac{1}{2})} $ representations of SL(2,$\mathbb{C}$) group. They obey
 $A(\omega)\sigma^{m}A^{\dagger}(\omega) =(e^{-\omega})^{m}_{\   n}\sigma^{n} $.

The commutation rules (\ref{42}) imply the following composition law for $N = 1$ Poincare supergroup:

\begin{align}\label{44}
g(\xi, \bar{\xi}, a, \omega)&g(\eta, \bar{\eta},b,\omega')=g(\Lambda, \bar{\Lambda}, c, \omega'')
\end{align}
where:

\begin{align}\label{45}	
&\Lambda^{\alpha} = \xi^{\alpha} + (A^{T}(\omega))^{\alpha}_{\   \beta}\eta^{\beta}\\ 
&\bar{\Lambda}^{\dot{\alpha}} = \bar{\xi}^{\dot{\alpha}} + (A^{\dagger})(\omega)^{\dot{\alpha}}_{\  \dot{\beta}}\bar{\eta}^{\dot{\beta}}\nonumber \\
&c^{m} = (e^{-\omega})^{m}_{\  n}b^{n} + a^{m} -i\xi^{\alpha}(\sigma^{m})_{\alpha \dot{\alpha}}(A^{\dagger})
(\omega)^{\dot{\alpha}}_{\   \dot{\beta}}\bar{\eta}^{\dot{\beta}}\nonumber \\
& -i\bar{\xi}^{\dot{\alpha}}(\sigma^{m})_{\alpha \dot{\alpha}}(A^{T})(\omega)^{\alpha}_{\   \beta}\eta^{\beta}\nonumber\\
&e^{\frac{i}{2}\omega''^{mn}M_{mn}} = e^{\frac{i}{2}\omega^{mn}M_{mn}}e^{\frac{i}{2}\omega'^{mn}M_{mn}}\nonumber
\end{align}

Now, dividing the $N=1$ Poincare supergroup by the Lorentz group one obtains $N=1$ Minkowski superspace. It is parametrized by four commuting real coordinates $x^{m}$, $m=0,1,2,3$ and four anticommuting grassmannian ones $\eta^{\alpha}$, $\bar{\eta}^{\dot{\alpha}}$, $\alpha$ = $1,2$, $\dot{\alpha}$=$\dot{1}$, $\dot{2}$ which obey the following condition: \\
\begin{align}
\label{46}
(\eta^{\alpha})^{\ast}=\bar{\eta}^{\dot{\alpha}}.
\end{align} 
The metric structure of bosonic sector is given by metric tensor  $g_{mn}$; in the fermionic sector the role of metric tensor is played by completely antisymmetric symbols (\ref{3}) and (\ref{4}).\\
The action of $N=1$ Poincare supergroup on Minkowski superspace implied by the above definitions of the superspace and the composition rule reads:   

\begin{align}\label{47}
&g(\xi, \bar{\xi}, a, \omega):(x^{m},\eta^{\alpha}, \bar{\eta}^{\dot{\alpha}})\longmapsto (x'^{m},\eta'^{\alpha}, \bar{\eta'}^{\dot{\alpha}})\\
&\eta'^{\alpha} = \xi^{\alpha} + (A^{T}(\omega))^{\alpha}_{\   \beta}\eta^{\beta}\nonumber \\ 
&\bar{\eta'}^{\dot{\alpha}} = \bar{\xi}^{\dot{\alpha}} + (A^{\dagger}(\omega))^{\dot{\alpha}}_{\  \dot{\beta}}\bar{\eta}^{\dot{\beta}} \nonumber \\
&x'^{m} = (e^{-\omega})^{m}_{\  n}x^{n} + a^{m} -i\xi^{\alpha}(\sigma^{m})_{\alpha \dot{\alpha}}(A^{\dagger}(\omega))^{\dot{\alpha}}_{\   \dot{\beta}}\bar{\eta}^{\dot{\beta}}\nonumber \\& -i\bar{\xi}^{\dot{\alpha}}(\sigma^{m})_{\alpha \dot{\alpha}}(A^{T}(\omega))^{\alpha}_{\   \beta}\eta^{\beta}\nonumber
\end{align}

In order to obtain deformed $N=1$ Minkowski superspace the commuting coordinates $x^{m}$, $\eta^{\alpha}$, $\bar{\eta}^{\dot{\alpha}}$ are replaced by noncommuting quantities $\hat{x^{m}}$, $\hat{\eta^{\alpha}}$, $\hat{\bar{\eta}}^{\dot{\alpha}}$ which satisfy some consistent commutation rules as well as relations given by eqs. (\ref{39}), (\ref{46}). If covariance of these rules is required the usual Poincare supertransformations have to be generalized somehow. This can be done starting with the assumptions (a-d) formulated above in the context of Euclidean case. Obviously, in the assumption c) the equation describing the action of generalized Euclidean transformations on deformed Euclidean superspace should be replaced by the following one:

\begin{align}\label{48}
&g(\hat{\xi,} \hat{\bar{\xi}}, \hat{a}, \hat{\omega}):(\hat{x}^{m},\hat{\eta}^{\alpha}, \hat{\bar}{\eta}^{\dot{\alpha}})
\longmapsto (\hat{x}'^{m},\hat{\eta}'^{\alpha}, \hat{\bar}{\eta'}^{\dot{\alpha}})\\
&\hat{\eta}'^{\alpha} = \hat{\xi}^{\alpha} + (A^{T}(\hat{\omega}))^{\alpha}_{\   \beta}\hat{\eta}^{\beta}\nonumber \\ 
&\hat{\bar{\eta}}'^{\dot{\alpha}} = \hat{\bar{\xi}}^{\dot{\alpha}} + 
(A^{\dagger}(\hat{\omega)})^{\dot{\alpha}}_{\  \dot{\beta}}\hat{\bar{\eta}}^{\dot{\beta}}\nonumber \\
&\hat{x}'^{m} = (e^{-\hat{\omega}})^{m}_{\  n}\hat{x}^{n} + \hat{a}^{m} -i\hat{\xi}^{\alpha}(\sigma^{m})_{\alpha \dot{\alpha}}
(A^{\dagger}(\hat{\omega})^{\dot{\alpha}}_{\   \dot{\beta}}\hat{\bar{\eta}}^{\dot{\beta}}\nonumber \\
& -i\hat{\bar{\xi}}^{\dot{\alpha}}(\sigma^{m})_{\alpha \dot{\alpha}}(A^{T}(\hat{\omega}))^{\alpha}_{\   \beta}\hat{\eta}^{\beta}.\nonumber
\end{align}

It gives the action of generalized Poincare transformations on deformed Minkowski superspace.\\

Now, let us assume the simplest deformation of grassmannian sector of Minkowski superspace :

\begin{align}\label{49}
\{\hat{\eta}^{\alpha}, \hat{\eta}^{\beta}\}=& C^{\alpha \beta}\\
\{\hat{\eta}^{\alpha},\hat{\bar{\eta}}^{{\dot{\beta}}}\}=&E^{\alpha \dot{\beta}} \nonumber \\ 
\{\hat{\bar{\eta}}^{\dot{\alpha}},\hat{\bar{\eta}}^{\dot{\beta}}\}=&\bar{C}^{\dot{\alpha} \dot{\beta}},\nonumber
\end{align}

The constants  $C^{\alpha \beta}$, $\bar{C}^{\dot{\alpha} \dot{\beta}}$ $E^{\alpha \dot{\beta}}$ must satisfy the constraints implied by
the counterpart of eq.(\ref{46}) for coordinates:

\begin{align*}
(C^{\alpha \beta})^{\ast}= \bar{C}^{\dot{\alpha} \dot{\beta}}\\
(E^{\alpha \dot{\beta}})^{\ast} = E^{\beta \dot{\alpha}}
\end{align*}

Then, proceeding as in the Euclidean case we get the remaining commutation rules which define Minkowski superspace deformations depending 
on two additional quantities $(\Pi^{m \alpha})^{\ast}$ = $\bar{\Pi}^{m \dot{\alpha}}$ and $\Theta^{mn}$
 
\begin{align}\label{50}
[\hat{x}^{m}, \hat{\eta}^{\alpha}]=&iE^{\alpha \dot{\rho}}(\sigma^{m})_{\rho \dot{\rho}}\hat{\eta}^{\rho} + 
iC^{\alpha \rho}(\sigma^{m})_{\rho \dot{\rho}}\hat{\bar{\eta}}^{\dot{\rho}}+ i\Pi^{m \alpha} \\
[\hat{x}^{m}, \hat{\bar{\eta}}^{\dot{\alpha}}]=&iE^{ \rho \dot{\alpha}}(\sigma^{m})_{\rho \dot{\rho}}\hat{\bar{\eta}}^{\dot{\rho}} + 
i\bar{C}^{{\dot{\alpha}} \dot{\rho}}(\sigma^{m})_{\rho \dot{\rho}}\hat{\eta}^{\rho}+ i\bar{\Pi}^{m \dot{\alpha}},\nonumber
\end{align} 

\begin{align}\label{51}
[\hat{x}^{m},\hat{x}^{n}] &= \frac{1}{2} \bar{C}^{\dot{\alpha} \dot{\beta}} (\sigma^{m})_{\gamma \dot{\alpha}}
(\sigma^{n})_{\delta \dot{\beta}}[\hat{\eta}^{\gamma}, \hat{\eta}^{\delta}] \\&+\frac{1}{2} C^{\alpha \beta} 
(\sigma^{m})_{\alpha \dot{\gamma}}(\sigma^{n})_{\beta \dot{\delta}}[\hat{\bar{\eta}}^{\dot{\gamma}}, \hat{\bar{\eta}}^{\dot{\delta}}]\nonumber \\
& + \frac{1}{2} E^{\alpha \dot{\beta}} ((\sigma^{m})_{\delta \dot{\beta}}(\sigma^{n})_{\alpha \dot{\gamma}} - 
(\sigma^{n})_{\delta \dot{\beta}}(\sigma^{m})_{\alpha \dot{\gamma}})[\hat{\eta}^{\delta}, \hat{\bar{\eta}}^{\dot{\gamma}}] \nonumber \\
&+ (\bar{\Pi}^{m \dot{\alpha}}(\sigma^{n})_{\sigma \dot{\alpha}} - \bar{\Pi}^{n \dot{\alpha}}
(\sigma^{m})_{\sigma \dot{\alpha}})\hat{\eta}^{\sigma} \nonumber \\
& + (\Pi^{m \alpha}(\sigma^{n})_{\alpha \dot{\sigma}} - \Pi^{n \alpha}(\sigma^{m})_{\alpha \dot{\sigma}})\hat{\bar{\eta}}^{\dot{\sigma}} +
 i\Theta^{mn},\nonumber
\end{align}

The superalgebraic sector of corresponding generalized Poincare transformations reads

\begin{align}\label{52}
[A^{T}(\hat{\omega}), \bullet] =&0\\
[A^{\dagger}(\hat{\omega}), \bullet] =&0\nonumber \\
[e^{-\hat{\omega}}, \bullet] =&0 \nonumber \\
\{\hat{\xi}^{\alpha}, \hat{\xi}^{\beta}\} =& C^{\alpha \beta}_{-} \nonumber \\
\{\hat{\bar{\xi}}^{\alpha}, \hat{\bar{\xi}}^{\beta}\} =& \bar{C}^{\dot{\alpha} \dot{\beta}}_{-}\nonumber \\ 
\{\hat{\xi}^{\alpha}, \hat{\bar{\xi}}^{\dot{\beta}}\} =& E^{\alpha \dot{\beta}}_{-}\nonumber\\ 
[\hat{a}^{m}, \hat{\xi}^{\alpha}]=&iE_{+}^{\alpha \dot{\rho}}(\sigma^{m})_{\rho \dot{\rho}}\hat{\xi}^{\rho} + 
iC_{+}^{\alpha \beta}(\sigma^{m})_{\rho \dot{\rho}}\hat{\bar{\xi}}^{\dot{\rho}}+ i\Pi^{m \alpha}_{-}\nonumber \\
[\hat{a}^{m}, \hat{\bar{\xi}}^{\dot{\alpha}}]=&iE^{ \rho \dot{\alpha}}_{+}(\sigma^{m})_{\rho \dot{\rho}}\hat{\bar{\xi}}^{\dot{\rho}} + 
i\bar{C}^{\dot{\alpha}\dot{\rho}}_{+}(\sigma^{m})_{\rho \dot{\rho}}\hat{\xi}^{\rho}+ i\bar{\Pi}^{m \dot{\alpha}}_{-}\nonumber \\
[\hat{a}^{m},\hat{a}^{n}] =& \frac{1}{2} \bar{C}_{-}^{\dot{\alpha} \dot{\beta}} (\sigma^{m})_{\gamma \dot{\alpha}}(\sigma^{n})_{\delta \dot{\beta}}
[\hat{\xi}^{\gamma}, \hat{\xi}^{\delta}] \nonumber \\
&+\frac{1}{2} C_{-}^{\alpha \beta} (\sigma^{m})_{\alpha \dot{\gamma}}(\sigma^{n})_{\beta \dot{\delta}}[\hat{\bar{\xi}}^{\dot{\gamma}},
 \hat{\bar{\xi}}^{\dot{\delta}}]\nonumber \\
& + \frac{1}{2} E_{-}^{\alpha \dot{\beta}} ((\sigma^{m})_{\gamma \dot{\beta}}(\sigma^{n})_{\alpha \dot{\delta}} - (\sigma^{n})_{\gamma \dot{\beta}}
(\sigma^{m})_{\alpha \dot{\delta}})[\hat{\xi}^{\gamma}, \hat{\bar{\xi}}^{\dot{\delta}}] \nonumber\\
&+ (\bar{\Pi}_{+}^{m \dot{\alpha}}(\sigma^{n})_{\sigma \dot{\alpha}} - \bar{\Pi}_{+}^{n \dot{\alpha}}
(\sigma^{m})_{\sigma \dot{\alpha}})\hat{\xi}^{\sigma}\nonumber \\
& + (\Pi_{+}^{m \alpha}(\sigma^{n})_{\alpha \dot{\sigma}} - \Pi^{n \alpha}_{+}(\sigma^{m})_{\alpha \dot{\sigma}})\hat{\bar{\xi}}^{\dot{\sigma}} +
 i\Theta^{mn}_{-}\nonumber
\end{align}

where we have defined\\

\begin{align}\label{53}
C^{\alpha \beta}_{\pm}&=[\delta^{\alpha}_{\gamma}\delta^{\beta}_{\delta}\pm(A^{T})^{\alpha}_{\  \gamma}(A^{T})^{\beta}_{\   \delta}]C^{\gamma \delta}\\
\bar{C}^{\dot{\alpha} \dot{\beta}}_{\pm}&=[\delta^{\dot{\alpha}}_{\dot{\gamma}}\delta^{\dot{\beta}}_{\dot{\delta}}
\pm(A^{\dagger})^{\dot{\alpha}}_{\ \dot{\gamma}}(A^{\dagger})^{\dot{\beta}}_{\   \dot{\delta}}]\bar{C}^{\dot{\gamma} \dot{\delta}}\nonumber \\
E^{\alpha \dot{\beta}}_{\pm}&=[\delta^{\alpha}_{\gamma}\delta^{\dot{\beta}}_{\dot{\delta}}
\pm((A^{T})^{\alpha}_{\   \gamma}A^{\dagger})^{\dot{\beta}}_{\  \dot{\delta}}]E^{\gamma \dot{\delta}}\nonumber \\ 
\bar{\Pi}^{m \dot{\alpha}}_{\pm}&=(\delta^{m}_{p}\delta^{\dot{\alpha}}_{\dot{\beta}} 
\pm(e^{-\hat{\omega}})^{m}_{p}(A^{\dagger})^{\dot{\alpha}}_{\   \dot{\beta}})\bar{\Pi}^{p \dot{\beta}}\nonumber\\
\Pi^{m \alpha}_{\pm}&=(\delta^{m}_{p}\delta^{\alpha}_{\beta} \pm(e^{-\hat{\omega}})^{m}_{\  \ p}(A^{T})^{\alpha}_{\    \beta})\Pi^{p \beta}\nonumber\\
\Theta^{mn}_{-}&= (\delta^{m}_{p}\delta^{n}_{q}-(e^{-\hat{\omega}})^{m}_{\  p}(e^{-\hat{\omega}})^{n}_{\  q})\Theta^{pq}.\nonumber
\end{align}

As already mentioned above, in the Minkowski superspace case, dotted and undotted grassmannian quantities are required to be related by conjugation transformation. This is additional, as compared to Euclidean case, condition determining the structure of both coordinates and parameters commutation rules. Due to it, in Minkowski superspace it is not possible to deform only commutation rules containing dotted quantities
(for instance) leaving the rules containing undotted ones unchanged.\\
Consistency of Jacobi identities and commutation rules of space-time coordinates ( as well as those of parameters ) implies some complicated constraints on $C^{\alpha \beta}$, $E^{\alpha \dot{\beta}}$  and $\Pi^{m \beta}$ constants, similarly as in Euclidean case. However, contrary to
the latter case there exist no matrices C and E with c-number elements which satisfy these constraints (see Appendix). In particular, there are no counterparts of "obvious" 
Euclidean solutions given by eqs. (\ref{30}),(\ref{31}). The elements of nontrivial C and E matrices have to be even elements of some Grassmann algebras. What is more
$\Pi^{m \beta}$ constants can not be arbitrary but they have to fulfill some additional constraint given by eqs:
  
\begin{align}
(\sigma^{n})_{\rho \dot{\rho}}(\bar{\Pi}^{m \dot{\rho}}\Pi^{p \rho}+ \Pi^{m \rho}\bar{\Pi}^{p \dot{\rho}}) -(\sigma^{m})_{\rho \dot{\rho}}(\bar{\Pi}^{n \dot{\rho}}\Pi^{p \rho}+ \Pi^{n \rho}\bar{\Pi}^{p \dot{\rho}})+cykl(p,m,n) =0.
\end{align} 

The superalgebraic sector of deformed Minkowski superspaces can be introduced in a similar way as in Euclidean case. It is described by 
eqs. (\ref{33}), (\ref{34}) and (\ref{35}) where the matrices $B^{\dagger}$ and $A^T$ are replaced by $A^{\dagger}$ and $A^{T}$, 
defined by eqs. (\ref{43}).

\section{Summary}
\label{pod}

In general, commutation rules describing superspace deformations (quantizations) are not compatible with the usual spacetime symmetries.
In the present paper, starting from the assumptions which are both simple and natural from "physical" point of view we presented a direct construction of generalized Euclidean/Poincare transformations which preserve a wide class of commutation rules defining deformations of the relevant superspaces (including most of those discussed in the literature). These generalized transformations act on deformed superspaces in the standard way. However, they depend on
noncommutative parameters which satisfy some consistent commutation relations. If the algebraic sector of transformations defined by these relations is consistent with coalgebraic structure one deals with quantum symmetry supergroup. It is the case for Euclidean superspace deformation given by eqs. (\ref{17}), (\ref{21}), (\ref{22}) and (\ref{26}), (\ref{27}) with structure constants given by eqs. (\ref{30}) or (\ref{31}). 
The algebraic sector of the corresponding quantum symmetry supergroup is given by eqs. (\ref{15}), (\ref{18}), (\ref{19}), (\ref{23}), (\ref{24}) and 
(\ref{28}), (\ref{29}) (with the same constants given by eqs. (\ref{30}) or(\ref{31})) while the coalgebraic one is described by 
eqs. (\ref{33}), (\ref{34}), (\ref{35}), (\ref{36}). This supersymmetric extension of $\theta$ - Euclidean group can be considered as global counterpart of appropriately twisted Euclidean superalgebra (which has been identified as nonanticommutative Hopf algebra preserving superspace deformation given by eqs. (\ref{30}) or (\ref{31})
(see Refs \cite{a55}, \cite{a56}, \cite{a57}, \cite{a58})). The intensively studied particular case of this deformation, the so called $N = \frac{1}{2}$ supersymmetry (see for instance Ref. \cite{a36}, \cite{a45}) corresponds to the solution $C \neq 0$ and $\Pi  = 0 = \Theta  $.\\
It is worth noting that there is no Minkowski counterpart of this Euclidean superspace deformation as long as elements of C, E matrices are assumed to
be c-numbers. This is basically due to the fact that in Minkowski superspace, unlike in Euclidean one, dotted and undotted grassmannian quantities are  related by conjugation transformation. One can avoid this sort of no-go result by allowing the elements of C, E matrices to belong to some  Grassmann algebra.

\section{Acknowledgment}

We thank Piotr Kosiński for fruitful discussion and reading the manuscript.\\
The paper is supported by the grant of he Ministry of Science N N202 331139 and Lodz University grant 506/1037.\\
The anonymous referee is acknowledged for comments which helped us to improve the paper.\\

Note added: recently there appeared the paper \cite{a67} dealing with the superextensions of some twist deformations of Minkowski
space-time.

\section{Appendix}

\subsection{Jacobi Identities}
\subsubsection{Euclidean superspace}

It is straightforward to check that the Jacobi identities are not satisfied automatically if C, D, E matrices entering coordinates or parameters
commutation relations are arbitrary. In fact, assuming the elements of these matrices to be c-numbers and imposing Jacobi identities results in
the following equations

\begin{align}
&(EXC)^{T}=-EXC\nonumber\\
&(DXE)^{T}=-DXE\nonumber \\
&EXE=-CXD\label{r3}
\end{align}

with symmetric C and D

\begin{align}\label{r4}
&C=C^{T} \ \mbox{(T - denotes transposition)} \\
&D=D^{T}. \nonumber
\end{align}

These equations should be satisfied by an arbitrary $2 \times  2$ matrix $X \equiv X_{n}\sigma^{n}$ , $X_{n} \in \mathbb{C} $.\\
Now, combining eqs. (\ref{r3}) written for $ X = E$ and for $ X = Id $ one arrives at contradiction if C, D, E matrices are assumed to be
nonzero ones. So, at least one of these matrices should be zero. Putting $ C = 0 $ or $ D = 0 $ one finds $ detE = 0 $. Taking then the
eigenvectors of matrix E (at least one of them is the zero eigenvector) as columns of matrix X implies that $ E = 0 $.\\
On the other hand if $ E = 0 $ than $ detC = 0 $ or $ detD = 0 $. Proceeding as in the previous case one finds that $ C = 0 $ or $ D = 0 $.\\
It can be directly checked that if $ C \neq 0 $ , $ D = 0 = E $ or $ D \neq 0 $ , $ C = 0 = E $ then the remaining Jacobi identities are satisfied 
without imposing any further constraints.\\
Finally, let us note that there exist Euclidean superspace deformations consistent with Jacobi identities and given by nonzero 
$C$, $D$, $E$, $\Pi$, $\Delta$ matrices provided the elements of these matrices belong to some Grassmann algebra.
(It follows from the relevant comutation rules that $C^{\alpha \beta}$, $D^{\dot{\alpha} \dot{\beta}}$, $E^{\alpha \dot{\beta}}$ should 
be even elements of this algebra while $\Pi^{m \alpha}$ and  $\Delta^{m \dot{\alpha}}$ the odd ones. An example of such matrices is given below: 

\begin{align*}
C^{\alpha\beta}=\begin{bmatrix}\xi^{1}\xi^{2}c^{1}&  &\xi^{1}\xi^{3}c^{2}\\ 
\xi^{1}\xi^{3}c^{2}&  &\xi^{1}\xi^{4}c^{3}\end{bmatrix}&  &c^{i}\in{\mathbb{C}},\;\ C^{\alpha\beta}=C^{\beta\alpha}\\ \\
D^{\dot{\alpha}\dot{\beta}}=\begin{bmatrix}\xi^{1}\xi^{5}d^{1}&  &\xi^{1}\xi^{6}d^{2}\\ \xi^{1}\xi^{6}d^{2}&  &\xi^{1}\xi^{7}d^{3}\end{bmatrix}
& &d^{i}\in \mathbb{C},\;\ D^{\dot{\alpha}\dot{\beta}}=D^{\dot{\beta}\dot{\alpha}}\\ \\
E^{\alpha\dot{\beta}}=\begin{bmatrix}\xi^{1}\xi^{8}e_{1}&  &\xi^{1}\xi^{9}e_{2}\\\xi^{1}\xi^{10}e_{3}&  &\xi^{1}\xi^{11}e_{4}\end{bmatrix}
& &e_{i}\in \mathbb{C}\\ \\
\Pi^{m\alpha}=\xi^{1}\tilde{\Pi}^{m\alpha},\;\  \tilde{\Pi}^{m\alpha}\in \mathbb{C}\\
\Delta^{m\dot{\alpha}}=\xi^{1}\tilde{\Delta}^{m\dot{\alpha}},\;\  \tilde{\Delta}^{m\dot{\alpha}}\in \mathbb{C}
\end{align*}
where:
\begin{align*}
&\{\xi^{A};\{\xi^{A},\xi^{B}\}=0,\  \ A,B=1...11\},
\end{align*}
 are generators of some Grassmann algebra.
 
\subsubsection{Minkowski superspace}

In Minkowski superspace case the counterpart of eqs. (\ref{r3}) reads:

\begin{align}
&(EXC)^{T}=-EXC\nonumber\\
&(CXE)^{T}=-\bar{C}XE \nonumber\\
&EXE=-CX\bar{C}\label{m3}
\end{align}

Inserting $ X= E $ into these equations leads to the conclusion that there are no nonzero E, C matrices satisfying eqs.(\ref{m3}) if the elements
of these matrices are assumed to be c-numbers. However, if these elements are allowed to belong to some Grassmann algebra then there exist 
nonzero E, C matrices defining Minkowski superspace deformation. For instance, such matrices can be taken in the form: 

\begin{align*}
C^{\alpha\beta}=\begin{bmatrix}c_{1}\xi^{1}(\xi^{1})^{\ast}\xi^{2}\xi^{3}&  &c_{2}\xi^{1}(\xi^{1})^{\ast}\xi^{4}\xi^{5}\\
c_{3}\xi^{1}(\xi^{1})^{\ast}\xi^{6}\xi^{7}&  &c_{4}\xi^{1}(\xi^{1})^{\ast}\xi^{8}\xi^{9}\end{bmatrix}\\
\end{align*}

\begin{align*}
\bar{C}^{\dot{\alpha}\dot{\beta}}=\begin{bmatrix}(c_{1})^{\ast}(\xi^{3})^{\ast}(\xi^{2})^{\ast}\xi^{1}(\xi^{1})^{\ast}
&  &(c_{2})^{\ast}(\xi^{5})^{\ast}(\xi^{4})^{\ast}\xi^{1}(\xi^{1})^{\ast}\\
(c_{3})^{\ast}(\xi^{7})^{\ast}(\xi^{6})^{\ast}\xi^{1}(\xi^{1})^{\ast}
&  &(c_{4})^{\ast}(\xi^{7})^{\ast}(\xi^{8})^{\ast}\xi^{1}(\xi^{1})^{\ast}\end{bmatrix}\\
\end{align*}

\begin{align*}
E^{\alpha\dot{\beta}}=\begin{bmatrix}e_{1}\xi^{1}(\xi^{1})^{\ast}\xi^{10}(\xi^{10})^{\ast}
&  &e_{2}\xi^{1}(\xi^{1})^{\ast}\xi^{11}(\xi^{11})^{\ast}\\
e_{3}\xi^{1}(\xi^{1})^{\ast}\xi^{11}(\xi^{11})^{\ast}&  &e_{4}\xi^{1}(\xi^{1})^{\ast}\xi^{12}(\xi^{12})^{\ast}\end{bmatrix}\\
\end{align*}

\begin{align*}
\Pi^{m\dot{\alpha}}=(\xi^{1})^{\ast}(\Pi^{m\alpha})^{\ast}
\end{align*}

where:

\begin{align*}
&\bar{C}^{\dot{\alpha}\dot{\beta}}=(C^{\alpha\beta})^{\ast}&
&(E^{\alpha\dot{\beta}})^{\ast}=E^{\beta\dot{\alpha}}\\
&\Pi^{m\alpha}=\xi^{1}\tilde{\Pi}^{m\alpha}&  & \tilde{\Pi}^{m\alpha}\in \mathbb{C}\\
\end{align*}

and:

\begin{align*}
&\xi^{A},(\xi^{A})^{\ast}:A=1...12,\ \ \{\xi^{A},\xi^{B}\}=0=\{\xi^{A},(\xi^{B})^{\ast}\}=0=\{(\xi^{A})^{\ast},(\xi^{B})^{\ast}\}\\
&((\xi^{A})^{\ast})^{\ast}=\xi^{A}\  \ (\xi^{A}\xi^{B})^{\ast}=(\xi^{B})^{\ast}(\xi^{A})^{\ast}=-(\xi^{A})^{\ast}(\xi^{B})^{\ast}
\end{align*}

are generators of some Grassmann algebra.


\newpage
\begin{thebibliography}{99}
\bibitem{a2}
S. Doplicher, K. Fredenhagen, J. E. Roberts, "The
Quantum Structure Of Space-Time At The Planck Scale
And Quantum Fields" Commun. Math. Phys. 172, 187
(1995): "Space-time quantization induced by classical
gravity," Phys. Lett. B 331, 39, (1994).
\bibitem{a3}
S. Majid, H. Ruegg, "Bicrossproduct structure of $\kappa$-Poincare group and noncommutative geometry", Phys. Lett. B334, 348 (1994)
\bibitem{a4}
S Zakrzewski,"Quantum Poincare group related to the $\kappa$-Poincare algebra", J. Phys. A: Math. Gen. 27 2075 (1994)
\bibitem{a5}
J. Madore, S.Schraml, P.Schupp, J. Wess "Gauge theory on noncommutative spaces", Eur. Phys. J. C16, 161 (2000) 
\bibitem{a6}
P. Aschieri, M. Dimitrijevic, F. Meyer and J. Wess, "Noncommutative
geometry and gravity", Class.Quant.Grav. 23 (2006) 1883-1912
\bibitem{a6.1}
P. Kosiński, J. Lukierski, P. Maślanka, "Local D = 4 field theory on $\Theta$-deformed
Minkowski space, Phys. Rev. D 62", 025004, (2000)
\bibitem{a7}
R. J. Szabo, "Quantum field theory on noncommutative spaces", Phys. Rept. 378,
207, (2003)
\bibitem{a8}
M. R. Douglas, N. A. Nekrasov, "Noncommutative field theory", Rev. Mod. Phys.
73, 977, (2001)
\bibitem{a8.1}
J. Wess, "Deformed coordinate space derivatives", arXiv:hep-th/0408080
\bibitem{a8.2}
J. Lukierski, M. Woronowicz, "New Lie-algebraic and quadratic deformations of
Minkovski space from twisted Poincare symmetries", Phys. Lett. B 633, 116–124, (2006) 
\bibitem{a9}
Letter of Heisenberg to Peierls (1930), in: Wolfgang Pauli, Scientific Correspondence, vol. II, 15, Ed. Karl von Meyenn, Springer-Verlag, (1985)
\bibitem{a10}
E. Witten, "Noncommutative geometry and string field theory", Nucl. Phys. B 268, 253 (1986).
\bibitem{a10.1}
M. R. Douglas, C. M. Hull, "D-branes and noncommutative torus", JHEP 9802 (1998) 008, hep-th/9711165
\bibitem{a10.2}
N. E. Mavromatos, R. J. Szabo, "Matrix D-brane dynamics, logarithmic operators and quantization of noncommutative spacetime", Phys. Rev. D 59 (1999)
104018 (hep-th/9808124)
\bibitem{a10.3}
C.S. Chu, P. M. Ho, "Noncommutative open string and D-brane", Nucl. Phys. B. 550 (1999), 151, (hep-th 9812219) 
\bibitem{a10.4}
C.S. Chu, P. M. Ho, "Constrained quantization of open string in background B field and noncommutative D-brane", Nucl. Phys. B. 568 (2000), 447, (hep-th 9906192)
\bibitem{a10.5}
V. Schomerus, "D-branes and deformation quantization", JHEP 9906 (1999) 030, hep-th/9908142 
\bibitem{a11}
N. Seiberg, E. Witten, "String theory and noncommutative geometry", JHEP 9909, 032, (1999).
\bibitem{a11.1}
F. Ardalan, H. Arfaei, M. M. Sheikh-Jabbari, "Dirac quantization of open strings and noncommutativity in branes", Nucl. Phys. B 576, 578, (2000)
\bibitem{a11.2}
F. Ardalan, H. Arfaei and M. M. Sheikh-Jabbari, "Noncommutative geometry from strings and branes", JHEP 9902, 016, (1999)
\bibitem{a16}
H.J. Groenewold, "On the Principles of elementary quantum mechanics", Physica 12, 405, (1946) 
\bibitem{a17}
J. E. Moyal, "Quantum mechanics as a statistical theory", Proc. Cambridge Phil. Soc. 45, 99 (1949).
\bibitem{a13}
M. Wilde, P. Lecomte, "Existence of star-products and of formal deformations of the Poisson Lie algebra of arbitrary symplectic manifolds", Lett. Math. Phys. 7, 487-496, (1983)
\bibitem{a15.1}
P. Kosiński, J. Lukierski, P. Maślanka, "Local field theory on $\kappa$-Minkowski space, star products and noncommutative translations", Czech.J.Phys. 50, 1283-1290, (2000)
\bibitem{a14}
L. Castellani, "Noncommutative geometry and physics: A review of selected recent results", Class. Quant. Grav. 17, 3377, (2000)
\bibitem{a15}
M. Kontsevich, "Deformation quantization of Poisson manifolds", Lett. Math. Phys. 66, 157-216, (2003)
\bibitem{a12}
A. Sykora, C. Jambor, "Realization of algebras with the help of star-products", arXiv:hep-th/0405268v1
\bibitem{a22}
A. Gerhold, J. Grimstrup, H. Grosse, L. Popp, M. Schweda, R.Wulkenhaar,
"The energy-momentum tensor on noncommutative spaces: Some pedagogical comments", arXiv:hep-th/0012112
\bibitem{a21}
A. Micu, M. M. Sheikh Jabbari, "Noncommutative $\Phi^4$ theory at two loops", JHEP 0101, 025, (2001)
\bibitem{a23}
T. Pengpan, X. Xiong, "A note on the non-commutative Wess-Zumino model", Phys. Rev. D 63, 085012 (2001)
\bibitem{a24}
M. Abou-Zeid, H. Dorn, "Comments on the energy-momentum ten-
sor in noncommutative field theories", Phys. Lett. B 514, 183 (2001)
\bibitem{a25}
L. Alvarez-Gaume, M. Vazquez-Mozo, "General properties of noncommutative field
theories", Nucl. Lett. B 668, 293, (2003) 
\bibitem{a26}
L. Alvarez-Gaume,J.l. Barbon, R. Zwicky, "Remarks on time space noncommutative field theories",
JHEP 0105, 057, (2001)
\bibitem{a27}
S.L. Woronowicz, Commun.Math. Phys. 111, 613,(1987) 
\bibitem{a28}
P.P.Kulish ed. Quantum groups, Lecture Notes in mathematics 1510 (Springer Verlag, 1992)
\bibitem{a28.1}
V.Chari, A Pressley "A guide to quantum groups", Cambridge University Press, (1994) 
\bibitem{a29}
A. Klimyk, K. Schmüdgen, "Quantum groups and their representations", Springer,  552, (1997)
\bibitem{a30}
S. Majid, "Foundations of quantum group theory", Cambridge University Press, 664, (2000)
\bibitem{a32}
R. Oeckl "Untwisting noncommutative $R^{d}$ and the equivalence of quantum field theories", Nucl. Phys.B 581, 559, (2000)
\bibitem{a32.1}
M. Chaichian, P. P. Kulish, K. Nishijima, A. Tureanu, "On a Lorentz-Invariant Interpretation of Noncommutative
Space-Time and Its Implications on Noncommutative QFT", Phys. Lett. B604, 98-102, (2004)
\bibitem{a31}
F. Koch, E. Tsouchnika, "Construction of $\Theta$-Poincare algebras and their invariants
on $M(\theta)$", Nucl. Phys. B717, 385, (2005) 
\bibitem{a33}
P. Aschieri, Ch. Blohmann, M. Dimitrijevic, F. Meyer, P. Schupp, J. Wess, "A gravity theory on noncommutative spaces", Class.  Quant. Grav. 22, 3511, (2005)
\bibitem{a34}
M. Chaichian, P. Presnajder, A. Tureanu, "New concept of relativistic invariance in NC
space-time: Twisted Poincare symmetry and its implication", Phys. Rev. Lett. 94, 151602, (2005)
\bibitem{a36}
P. P. Kulish "Twists of quantum groups and noncommutative field theory", arXiv:hep-th/0606056v1
\bibitem{a37}
V.G. Drinfeld, "Quasi-Hopf algebras", Leningrad Math. J. 1, 1419–1457, (1990)
\bibitem{a38}
N.Yu. Reshetikhin, L.A. Takhtajan, L.D. Faddeev, "Quantization of Lie groups and
Lie algebras", Algebra and Analysis 1 (1989) 178–206
\bibitem{a40}
P.Kosiński, P. Maślanka, "Lorentz-Invariant Interpretation of Noncommutative Space-Time - global version", arXiv:hep-th/0408100v3
\bibitem{a45}
N. Seiberg, "Noncommutative superspace, N=1/2 supersymmetry, field theory and string theory", JHEP 0306, 010, (2003)
\bibitem{a46}
N. Berkovits, N. Seiberg, "Superstrings in graviphoton background and N = 1/2 + 3/2 supersymmetry",
JHEP 0307, 010 (2003)
\bibitem{a47}
P. Kosiński, J. Lukierski, P. Maślanka, J. Sobczyk, "Quantum deformation of the Poincare supergroup and $\kappa$ deformed superspace", J. Phys. A
27, (1994),6827  
\bibitem{a48}
P. Kosiński, J. Lukierski, P. Maślanka, "Quantum deformations of space-time SUSY and noncommutative superfield
theory", arXiv:hep-th/0011053.
\bibitem{a50}
D. Klemm, S. Penati, L. Tamassia, "Non(anti)commutative superspace", Class. Quant.
Grav. 20, 2905, (2003) 
\bibitem{a51}
S. Terashima, J. T. Yee, "Comments on noncommutative superspace", JHEP 0312, 053,
(2003) 
\bibitem{a52}
S. Ferrara, M. A. Lledo, O. Macia, "Supersymmetry in non-commutative superspaces", JHEP 0309, 068, (2003)
\bibitem{a53}
V. Nazaryan, C. Carlson, "Field theory in noncommutative Minkowski superspace", Phys.Rev. D71, 025019, (2005) 
\bibitem{a54}
M. Arai, M. Chaichian, K. Nishijima , A. Tureanu, "Nonanticommutative supersymmetric field theory and quantum shift", arXiv:hep-th/0604029
\bibitem{a55}
Y. Kobayashi, S. Sasaki, "Lorentz invariant and supersymmetric interpretation of
noncommutative quantum field theory", Int. J. Mod. Phys. A 20, 7175-7188, (2005) 
\bibitem{a56}
B. M. Zupnik, "Twist-deformed supersymmetries in nonanticommutative superspaces", Phys.
Lett. B 627, 208, (2005) 
\bibitem{a57}
M. Ihl, C. Saemann, "Drinfeld-twisted supersymmetry and nonanticommutative superspace",
JHEP 0601, 065, (2006) 
\bibitem{a58}
M. Irisawa, Y. Kobayashi, S. Sasaki, "Drinfeld twisted superconformal algebra and structure
of unbroken symmetries", Prog. Theor. Phys. 118, 83-96, (2007) 
\bibitem{a59}
Y. Kobayashi, S. Sasaki, "Non-local Wess-Zumino model on nilpotent noncommutative
superspace", Phys. Rev. D72, 065015, (2005) 
\bibitem{a60}
B. A. Qureshi, "Twisted supersymmetry, fermion-boson mixing and removal of UV-IR mixing", arXiv:hep-th/0602040
\bibitem{a61}
M. Dimitirijevic, V. Radovanovic, J. Wess "Field theory on nonanticommutative superspace", JHEP 0712, 059, (2007)
\bibitem{a62}
Yu. I. Manin, Commun. Math. Phys. 123, 163, (1989)
\bibitem{a63}
L.A Takhtajan, Adv. Studies Pure Math. 19, 435, (1989)
\bibitem{a64}
E.Corrigan, D.B. Fairlie, P.Fletcher, R.Sasaki, J. Math. Phys. 31, 776, (1990) 
\bibitem{a65}
J.Wess, B. Zumino, Nucl. Phys. B (Proc. suppl), 18B, 302, (1990) 
\bibitem{a66}
C. Gonera, M. Wodzisławski, to be submitted
\bibitem{a41}
J. Wess, Jonathan Bagger, "Supersymmetry and Supergravity", Princeton University Press   272  (1991)
\bibitem{a42}
J. D. Lykken, "Introduction to Supersymmetry", arXiv:hep-th/9612114v1
\bibitem{a43}
P. C. West, "Introduction to Supersymmetry and Supergravity", World Scientific Publishing Company 289 (1986)
\bibitem{a44}
P. C. Argyres, "Introduction to supersymmetry", http://www.physics.uc.edu/ argyres/661/index.html. 
\bibitem{a67}
A.Borowiec, J. Lukierski, M. Mozrzymas, V.N. Tolstoy  "$N = \frac{1}{2}$ Deformations of Chiral Superspaces
from New Twisted Poincare and Euclidean Superalgebras", arXiv:1112.1936v1 
\end{thebibliography}
\end{document}